\documentclass[preprint,12pt,review]{elsarticle}
\usepackage[utf8]{inputenc}
\usepackage{graphicx}
\usepackage{amssymb}
\usepackage{soul}
\usepackage{appendix}
\usepackage{amsthm}
\usepackage{setspace}
\usepackage{amsmath}
\makeatletter

\renewcommand\tagform@[1]{\maketag@@@{\ignorespaces#1\unskip\@@italiccorr}}
\renewcommand\theequation{(\oldtheequation)}
\makeatother
\usepackage{amssymb}
\usepackage{textcomp}
\usepackage{gensymb}
\usepackage{xspace}
\usepackage{longtable}
\usepackage{booktabs}
\usepackage[normalem]{ulem}
\usepackage[version=3]{mhchem} 
\usepackage[usenames]{color}
\usepackage[T1]{fontenc}
\usepackage{xr-hyper} 
\usepackage[pdfpagelabels,bookmarks]{hyperref}
\usepackage{upgreek}
\usepackage{siunitx}
\usepackage{caption}
\usepackage{pifont}
\usepackage[english]{babel}
\usepackage{geometry} 
\biboptions{sort&compress,comma}

\addto\extrasenglish{

}
\renewcommand{\eqref}[1]{Eq.~\ref{#1}}

\newcommand{\eqsref}[2]{Eqs.~\ref{#1}~\&~\ref{#2}}

\newcommand\unappendix{\par
  \setcounter{section}{0}%
  \setcounter{subsection}{0}%
  \gdef\thesection{\@arabic\c@section}}
  \renewcommand{\appendixautorefname}{Section}
\makeatother

\usepackage{newfloat}
\DeclareFloatingEnvironment[
  fileext   = los,                      
  listname  = {List of Files},          
  name      = File,                     
  placement = htp                       
]{file}

\begin{document}

\begin{frontmatter}

\title{Homogenized Lattice Boltzmann Model for Simulating Multi-Phase Flows in Heterogeneous Porous Media}

 \author[DLR,HIU]{Martin P. Lautenschlaeger}\corref{1}
 \cortext[1]{Corresponding author:}
 \ead{Martin.Lautenschlaeger@dlr.de}
 \author[DLR,HIU]{Julius Weinmiller}
 \author[DLR,HIU]{Benjamin Kellers}
 \author[DLR,HIU]{Timo Danner}
 \author[DLR,HIU,UUlmE]{Arnulf Latz}
 
 \address[DLR]{German Aerospace Center (DLR), Institute of Engineering Thermodynamics, 70569 Stuttgart, Germany}
 \address[HIU]{Helmholtz Institute Ulm for Electrochemical Energy Storage (HIU), 89081 Ulm, Germany}
 \address[UUlmE]{Ulm University (UUlm), Institute of Electrochemistry, 89081 Ulm, Germany}

\begin{abstract}

A new homogenization approach for the simulation of multi-phase flows in heterogeneous porous media is presented. It is based on the lattice Boltzmann method and combines the grayscale with the multi-component Shan-Chen method. Thus, it mimics fluid-fluid and solid-fluid interactions also within pores that are smaller than the numerical discretization. The model is successfully tested for a broad variety of single- and two-phase flow problems. Additionally, its application to multi-scale and multi-phase flow problems in porous media is demonstrated using the electrolyte filling process of realistic 3D lithium-ion battery electrode microstructures as an example. The new method shows advantages over comparable methods from literature. The interfacial tension and wetting conditions are independent and not affected by the homogenization. Moreover, all physical properties studied here are continuous even across interfaces of porous media. The method is consistent with the original multi-component Shan-Chen method. It is accurate, efficient, easy to implement, and can be applied to many research fields, especially where multi-phase fluid flow occurs in heterogeneous and multi-scale porous media.

\end{abstract}


\end{frontmatter}

\journal{arXiv}



\section{Introduction}
Fluid flow in porous media plays an important role in many technical and natural processes such as hydrogeology, reservoir and process engineering, electrochemical energy storage, or medical applications. Most of these examples involve complex flow phenomena such as transport of solutes, reactions, or the interaction of multiple phases or immiscible fluid components \cite{Kang2007,Dentz2011,Steefel2005,Baveye2017,Laubach2019,Yuan2019}, structures that are heterogeneous regarding their chemical composition and wetting properties \cite{Dentz2011,Blunt2013,Laubach2019,Zhang2020}, and pore sizes that range from nanometers to the macroscale \cite{Sok2010,Bai2013,Blunt2013,Zhang2016,Kang2019,Soulaine2019,Zhang2020,Mehmani2020}. Thus, and because most of the interesting physical phenomena happen on the pore scale, they are hard to study experimentally \cite{Sok2010,Kang2002,Dentz2011,Mehmani2020}.

Therefore, in the literature often direct numerical simulations and more specifically the lattice Boltzmann method (LBM) are used to conduct pore-scale simulations. LBM is a reliable tool for studying multi-scale and multi-physics transport processes within complex porous geometries \cite{Krueger2016,Liu2016}. It has also been successfully applied to solve multi-phase flows in high-resolution real-world image data of porous media samples that were recorded using X‐ray micro‐computed tomography ($\mu$-CT) or focused ion beam scanning electron microscopy (FIB-SEM) \cite{Sok2010,Blunt2013,Chen2014,Liu2016,Zhang2016,Baveye2017,Kang2019,Zhang2020,Mehmani2020}. 

Unfortunately, LBM is computationally expensive, especially when simultaneously simulating flow in structurally resolved pores at different length scales. Therefore, homogenization methods have been developed, where the detailed structure of pores at the smallest length scale is ignored and, instead, the flow is described by a Darcy-Brinkman-type approach. A volume average of the structurally resolved geometry is taken and its effects on fluid flow are mimicked as permeability-related parameter. These homogenization methods can be basically subdivided into two groups. Those are the Brinkman force-adjusted models (BF) \cite{Spaid1997,Freed1998,Guo2002,Kang2002,Ginzburg2008,Gao2014,Ginzburg2015,Kang2019}, where a drag force is applied locally, and the grayscale models (GS) \cite{Dardis1998,Thorne2004,Chen2008,Walsh2009,Zhu2013,Yoshida2014,Yehya2015,Ginzburg2015,Ginzburg2016,Zhu2018}, where flow populations are partially bounced back to mimic flow resistance. 

Although the aforementioned homogenization methods have been heavily discussed and further developed for single-phase fluids \cite{Chen2008,Walsh2009,Zhu2013,Ginzburg2015,Ginzburg2016,Zhu2018}, this is not the case for multi-phase or multi-component fluids. Only a few methods combining GS and multi-phase physics \cite{McDonald2016,Zalzale2016,Lei2019} as well as methods combining other homogenization approaches with multi-component physics \cite{Ning2019,An2020} have been reported recently. However, despite the fact that the multi-component Shan-Chen method (MCSC) is most widely used for studying all kinds of immiscible fluids, only one homogenized method has been developed combining GS with MCSC \cite{Pereira2016}. This method is however not fully consistent with the original MCSC and shows deficiencies with respect to heterogeneous porous media.

Therefore, in the current paper, a new model is presented, that follows the approach of Pereira \cite{Pereira2016} and combines GS by Walsh~\textit{et al.} (GS-WBS) \cite{Walsh2009} with MCSC \cite{Shan1993}. It is therefore called the homogenized multi-component Shan-Chen method (HMCSC) in the following. GS-WBS is chosen as it is known to recover Darcy–Brinkman flow, conserves mass, and allows an efficient computational parallelization as only local bounce-back operations are performed. MCSC uses a physically-based approach to model fluid-fluid and solid-fluid interactions without the need for interface tracking. It also achieves a good compromise between computational efficiency and physical reality, and thus is widely adopted for modeling immiscible fluids.

The HMCSC inherits all positive features from the aforementioned models, but overcomes their deficiencies which are mainly related to the discontinuity of properties in heterogeneous porous media. For example, using HMCSC, the interfacial tension and the wetting properties are constant and not affected by the homogenization. Thus, especially the MCSC-related model parameters can be chosen consistently to the original MCSC and no further parametrization is required. Besides, the new method switches freely between free-flow and Darcy regime, and can be also applied to study single-phase flows. 

As part of this paper, the HMCSC was rigorously tested for a broad variety of single-phase and two-phase flow benchmark cases that are relevant in the context of porous media. Those were Stokes-Brinkman-Darcy flow under Couette and Poiseuille conditions, fluid flow in stratified heterogeneous porous media and partially porous channels, as well as steady bubble tests and Washburn-type capillary flow. It was also shown to predict Buckley-Leverett waterflooding to some extent (cf.\ \autoref{sec:BuckleyLeverett} in the Supporting Information). The results were compared with analytical and semi-analytical solutions where available. Finally, the HMCSC was applied to a two-phase flow issue of current research interest in the field of electrochemical energy storage: The electrolyte filling of lithium-ion battery microstructures with partially permeable nanoporous components. However, other research fields where multi-phase fluid flow occurs in multi-scale porous media can benefit from the new method, too. In the context of hydrology, geoscience and petroleum engineering, potential applications are the prediction of microbiologically affected groundwater flow \cite{Ghezzehei2012,Hassannayebi2021}, geologic carbon storage or sequestration \cite{Krevor2012,Mehmani2018}, and the recovery of oil, dry natural gas, or shale gas from tight gas sandstones \cite{Mehmani2015,Mehmani2020}, carbonates \cite{Mehmani2015,Mehmani2020} and shale formations \cite{Soulaine2019}, respectively.

This paper is organized as follows. In \autoref{sec:Model}, the HMCSC is described. In \autoref{sec:Validation}, it is tested for a broad variety of benchmark cases and the corresponding results and features of the method are discussed. In \autoref{sec:Results}, the results of the electrolyte filling simulations in realistic and partially homogenized battery microstructures are presented. Finally, conclusions are drawn in \autoref{sec:Conclusion}.

\section{Model}  \label{sec:Model}
The LBM fundamentals, including the determination of macroscopic variables, as well as the underlying methods are given in \autoref{sec:Appendix_LBM}. For further information, especially regarding GS-WBS and MCSC, the reader is directed to the corresponding references \cite{Walsh2009,Shan1993}. In the following, only relevant parts that are necessary to understand the HMCSC are described. The full declaration of notations is also given in \autoref{sec:Appendix_LBM}.

The main equation of the HMCSC that combines the GS-WBS homogenization approach \cite{Walsh2009} with the MCSC and the Shan-Chen forcing scheme \cite{Shan1993} is
\begin{align}
    \label{eq:LGS}
    \begin{split}
    f_{i,\sigma}\left(\mathbf{x}+\mathbf{c}_i \Delta t, t + \Delta t \right) =&\quad\  (1-n_{\mathrm{s},\sigma}(\mathbf{x}))f_{i,\sigma}\left(\mathbf{x}, t\right) \\\
    &- (1-n_{\mathrm{s},\sigma}(\mathbf{x}))\frac{\Delta t}{\tau_{\sigma}} \left(f_{i,\sigma}\left(\mathbf{x}, t\right) - f^{\mathrm{eq}}_{i,\sigma}\left(\mathbf{x}, t\right)\right) \\\
    &+ n_{\mathrm{s},\sigma}(\mathbf{x}) f_{\bar{i},\sigma}\left(\mathbf{x}, t^{\ast}\right).
    \end{split}
\end{align}
Here, $\boldsymbol{f}_{\sigma}$ is the distribution function of the component $\sigma$, $\boldsymbol{f}^\mathrm{eq}_{\sigma}$ is the Maxwell-Boltzmann equilibrium distribution function (cf. \autoref{eq:Maxwell} in the Appendix), and $\tau_{\sigma}$ is the relaxation time. The last term in \autoref{eq:LGS} corresponds to the bounce-back scheme. The parameter $n_{\mathrm{s},\sigma}$ comes from the homogenization approach. Originally it was called the solid fraction which is due to the intuitive interpretation of relating flow properties to the solid volume fraction. In the following, the same parameter is called the bounce-back fraction to highlight its technical origin and to prevent a misinterpretation. Note that the bounce-back fraction has to be chosen to retrieve the permeability and should be seen as a space-dependent, internal model parameter which is not necessarily directly proportional to the amount of solid material in a lattice cell. Its general relation to the permeability for a single-phase fluid \cite{Walsh2009} is
\begin{align}    
    \label{eq:GS_perm}
    k = \frac{1-n_\mathrm{s}}{2 n_\mathrm{s}} \nu \Delta t.
\end{align}

Similar to the MCSC \cite{Shan1993}, \autoref{eq:LGS} is solved for each component $\sigma$ involved in the multi-phase flow. Thus, all lattice cells are occupied by every component simultaneously. A cell belonging to component $\sigma$ is composed of the main component density $\rho_{\sigma}$ and the dissolved densities with $\rho_{\text{dis}} \ll \rho_{\sigma}$. The fluid-fluid and solid-fluid interactions are incorporated as interaction forces via $\boldsymbol{f}_\sigma^\mathrm{eq}$ (cf.\ \autoref{eq:SC_u_eq}). They are physically motivated by a pseudopotential that similar to molecular dynamic simulations models molecular interactions to recover cohesion and adhesion, e.g., in wetting or transport processes \cite{Lautenschlaeger2019,Lautenschlaeger2019a,Diewald2020,Lautenschlaeger2020,Schmitt2022}.

The separation of component $\sigma$ by another component $\bar{\sigma}$ is driven by a fluid-fluid interaction force $\boldsymbol{F}_\mathrm{inter}$
\begin{align}
    \label{eq:SC_F_inter}
    \boldsymbol{F}_\mathrm{inter,\sigma}\left(\mathbf{x}\right) = -\rho_{\sigma}\left(\mathbf{x}\right) G_{\mathrm{inter},\sigma\bar{\sigma}} \sum_{i}^{}w_i \rho_{\bar{\sigma}}\left(\mathbf{x}+\mathbf{c}_i \Delta t\right)\mathbf{c}_i \Delta t,
\end{align}
where the interaction parameter $G_{\mathrm{inter},\sigma\bar{\sigma}}$ determines the strength of the cohesion.

The wettability or adhesion of the component $\sigma$ at a solid wall is modeled with the solid-fluid interaction force $\boldsymbol{F}_\mathrm{ads,\sigma}$
\begin{align}
    \label{eq:SC_F_ads}
    \boldsymbol{F}_\mathrm{ads,\sigma}\left(\mathbf{x}\right) = -\rho_{\sigma}\left(\mathbf{x}\right) G_\mathrm{ads,\sigma} \sum_{i}^{}w_i s\left(\mathbf{x}+\mathbf{c}_i \Delta t\right)\mathbf{c}_i \Delta t,
\end{align}
where the adhesion parameter $G_\mathrm{ads,\sigma}$ determines the wetting behavior and $s$ is an indicator function which is $s = n_{\mathrm{s},\sigma}$ here. 

It was shown by Huang~\textit{et al.} \cite{Huang2007} how $G_{\mathrm{inter},\sigma\bar{\sigma}}$ and $G_\mathrm{ads,\sigma}$ relate to the interfacial tension $\gamma$ and the contact angle $\theta$, respectively. It will be shown in \autoref{sec:Validation}, that the same parametrization can be used for the HMCSC, too, which underlines its physical consistency with the original MCSC. 

Additional external forces $\boldsymbol{F}_\mathrm{ext}$ which act on all components, such as gravity, are distributed to each component $\sigma$ by their density ratios 
\begin{align}    
    \label{eq:SC_F_ext}
    \boldsymbol{F}_\mathrm{ext,\sigma} = \frac{\rho_{\sigma}}{\rho} \boldsymbol{F}_\mathrm{ext},
\end{align}
where $\rho=\sum_{\sigma}^{}\rho_{\sigma}$ is the total density of all components in a lattice cell.

All aforementioned force contributions are summarized to the total force $\boldsymbol{F}_\mathrm{tot,\sigma}=\boldsymbol{F}_\mathrm{inter,\sigma}+\boldsymbol{F}_\mathrm{ads,\sigma}+\boldsymbol{F}_\mathrm{ext,\sigma}$. Using the Shan-Chen forcing approach, $\boldsymbol{F}_\mathrm{tot,\sigma}$ is finally incorporated into $\boldsymbol{f}_\sigma^\mathrm{eq}(\rho_\sigma, \textbf{u}^{\mathrm{eq}}_\sigma)$ (cf. \autoref{eq:Maxwell} in the Appendix) as a force-induced equilibrium velocity shift of each component
\begin{align}    
    \label{eq:SC_u_eq}
    \textbf{u}^{\mathrm{eq}}_{\sigma} = \frac{\sum_{\sigma}^{}\rho_{\sigma}\mathbf{u}_{\sigma}/\tau_{\sigma}}{\sum_{\sigma}^{}\rho_{\sigma}/\tau_{\sigma}} + \frac{\tau_{\sigma}\boldsymbol{F}_\mathrm{tot,\sigma}}{\rho_{\sigma}}.
\end{align}

Note that the equilibrium velocity $\textbf{u}^{\mathrm{eq}}$ must not be confused with the macroscopic streaming velocity of the mixture. For the HMCSC, the latter is given by
\begin{align}    
    \label{eq:HMCSC_u_macro}
    \textbf{u}_\mathrm{macro} = \frac{1}{\rho} \sum_{\sigma}^{}(1-n_{\mathrm{s},\sigma})\left(\sum_{i}^{} f_{i,\sigma} \mathbf{c}_i + \frac{\boldsymbol{F}_\mathrm{tot,\sigma} \Delta t}{2} \right),
\end{align}
where the factor $(1-n_{\mathrm{s},\sigma})$ comes from the homogenization approach \cite{Walsh2009,Yehya2015,Pereira2016}.

\section{Model Validation} \label{sec:Validation} 
The HMCSC has been implemented in the open-source LBM tool \textit{Palabos} (version 2.3) \cite{Latt2021}. This extended version of \textit{Palabos} was used to test the new model for typical porous media benchmark scenarios for single- and two-phase flow. The benchmark scenarios were chosen to cover a wide range of levels of complexity and are discussed in the following. All results are given in lattice units or dimensionless units and compared with analytical or semi-analytical solutions where available. Only 2D simulations were conducted for the validation. A 3D application of the HMCSC is described in \autoref{sec:Results}. All relevant model parameters are given in \autoref{tab:Parameters}. Unless specified otherwise, they were used for all simulations of the present work. For studying single-phase flows using the HMCSC, the MCSC-related model parameters were set to zero, i.e.\ $G_{\mathrm{inter},\sigma\bar{\sigma}}=G_{\mathrm{ads},\sigma}=0.0$. Under these conditions, the model reduces to the GS-WBS with Shan-Chen forcing scheme and no-slip boundary conditions at solids. 
 
 \begin{table}[!ht]
	\centering
	\caption{Overview of the physical quantities of the system consisting of fluid 1 and fluid 2. Values are given in LBM units (lu: length unit; ts: time step; mu: mass unit).}
	\begin{tabular}{l l}
		\toprule \toprule
        density & $\rho_1=\rho_2=0.99$ $\frac{\text{mu}}{\text{lu³}}$ \\ 
                & $\rho_{\text{dis,1}}=\rho_{\text{dis,2}}=0.01$ $\frac{\text{mu}}{\text{lu³}}$ \\
        kinematic viscosity & $\nu_{\mathrm{1}}=\nu_{\mathrm{2}}=1.667\cdot10^{-1}$ $\frac{\text{lu²}}{\text{ts}}$ \\
        surface tension & $\gamma=7.68\cdot10^{-2}$  $\frac{\text{mu}}{\text{ts²}}$ \\ \hline
        model parameters & $\tau_1=\tau_2=1.0$ \\
        & $n_\mathrm{s}=n_{\mathrm{s},1}=n_{\mathrm{s},2}$ \\
        & $G_\mathrm{inter,12}=G_\mathrm{inter,21}=1.75$ \\
        & $G_\mathrm{ads,2}=-G_\mathrm{ads,1}=\frac{1}{4}G_\mathrm{inter,12}(\rho_1-\rho_{\mathrm{dis,2}})\cos\theta$ \cite{Huang2007}\\
        & $\Delta \mathbf{x}=\Delta t = 1$ \\
		\bottomrule \bottomrule 
	\end{tabular}%
	\label{tab:Parameters}%
\end{table}%

\subsection{Permeability} \label{sec:Permeability}
The key parameter of the homogenization is the bounce-back fraction $n_\mathrm{s}$. For GS-WBS it was shown to be related to the permeability $k$ following \autoref{eq:GS_perm} \cite{Walsh2009}. It is shown in the following that \autoref{eq:GS_perm} is also true when using the HMCSC for single-phase flows. 

The simulation domain represents a fully homogenized porous medium, where $n_\mathrm{s}$ was identical in all lattice cells. The system was fully periodic, and its dimensions along the $x$- and $y$-direction were $H=50$\,lu each. The fluid flow was driven by the body force $F_\mathrm{ext}=10^{-5}$\,(mu\,lu)/ts² in $+x$-direction.
 
The permeability $k$ was calculated using Darcy's law for a single-component fluid
\begin{align}    
    \label{eq:Permeability}
    k = \frac{\nu\langle\rho u\rangle}{F_\mathrm{ext}},
\end{align}
where $\langle\rho u\rangle$ is the average momentum in $x$-direction. 

In addition, the relative permeabilities $k_{\mathrm{r},\sigma}$ were determined for a two-phase flow in the absence of adhesion $\left(G_{\mathrm{ads},\sigma}=0.0\right)$
\begin{align}    
    \label{eq:relativePermeability}
    k_{\mathrm{r},\sigma} = \frac{k(S_\sigma)}{k(S_\sigma=1.0)}.
\end{align}
Here, the index $\sigma$ denotes fluid 1 or fluid 2, respectively. $S_\sigma$ is the saturation of the simulation domain with fluid $\sigma$, and the right-hand side is the ratio between the permeability at a certain saturation $S_\sigma$ with the corresponding single-phase permeability, i.e.\ $S_\sigma=1.0$, both determined using \autoref{eq:Permeability}.

For the single-phase flow, $n_\mathrm{s}$ was varied in the range  $n_\mathrm{s}=[0.0001,0.9999]$. For the two-phase flow, exemplary values $\left(n_\mathrm{s}=\{0.1,0.5,0.9\}\right)$ were chosen, where for each value of $n_\mathrm{s}$, $S_\sigma$ was varied in the range $S_{\mathrm{1}}=1-S_{\mathrm{2}}=[0,1]$. 

\begin{figure}
	\centering
	  \includegraphics[width=1.0\textwidth]{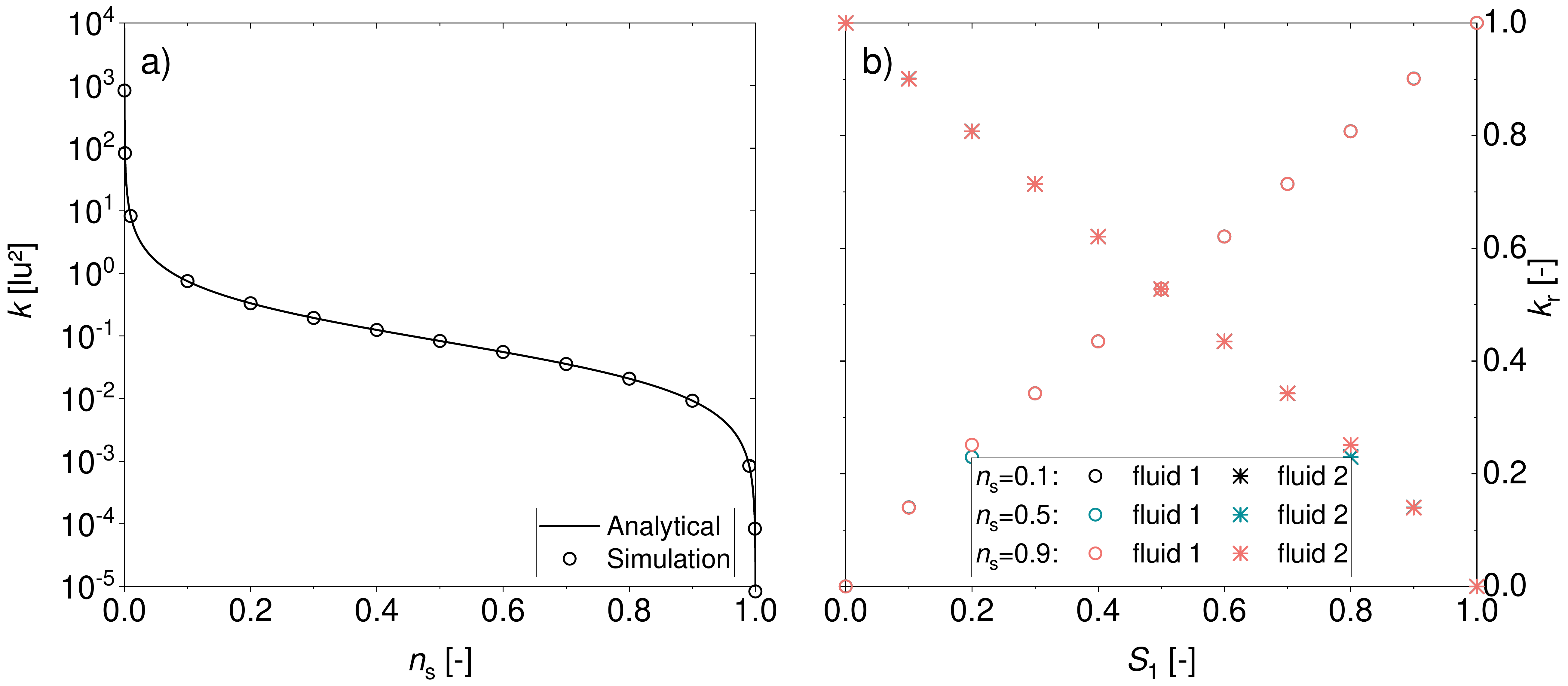}
	\caption{Profiles of a) the permeability $k$ for single-phase flow and b) the relative permeabilities $k_{\mathrm{r}}$ for two-phase flow as a function of $n_\mathrm{s}$ and the saturation $S_{\mathrm{1}}$. The simulation results are denoted by the symbols. For $k$ also the analytical solution (solid line, cf.\ \autoref{eq:GS_perm}) is shown. }
	\label{fig:Permeability}
\end{figure}

\autoref{fig:Permeability}a) shows the simulation results and the analytical solution (cf.\ \autoref{eq:GS_perm}) of $k$ for the single-phase flow. They are in excellent agreement which confirms the consistency with the GS-WBS and the applicability of \autoref{eq:GS_perm} for the HMCSC. \autoref{fig:Permeability}b) shows the simulation results of $k_{\mathrm{r},1}$ and $k_{\mathrm{r},2}$ for different $n_\mathrm{s}$ and as a function of $S_{\mathrm{1}}$. Hardly any dependence on $n_\mathrm{s}$ was observed. The results follow an almost linear trend which is typical for fluids with low interfacial tension \cite{Mu2019}. Moreover, as both fluids have identical properties (cf. \autoref{tab:Parameters}), it holds $k_{\mathrm{r},1}(S_{\mathrm{1}})=k_{\mathrm{r},2}(1-S_{\mathrm{1}})$.

\subsection{Darcy-Brinkman Flow} \label{sec:DarcyBrinkman}
A critical requirement for all homogenization approaches is that they recover Darcy-Brinkman-type flow behavior in porous media \cite{Chen2008,Zhu2013,Yehya2015,Ginzburg2016,Pereira2016,Zhu2018,Ning2019}, which is described by the Darcy-Brinkman equation
\begin{align}    
    \label{eq:DarcyBrinkmanEq}
    \frac{\nu_\mathrm{B}}{\phi} \frac{\partial^2 u}{\partial x^2} + F_\mathrm{ext} - \frac{\nu}{k}u = 0.
\end{align}
Here, $\nu_\mathrm{B}$ is the effective Brinkman viscosity, $\nu$ is the viscosity of the fluid, $F_\mathrm{ext}$ is the driving force, and $\phi$ and $k$ are the porosity and permeability of the porous medium, respectively.

The analytical solution of \autoref{eq:DarcyBrinkmanEq} depends on the choice of boundary conditions. Three Darcy-Brinkman flow types were studied and compared to their analytical solutions. The different variants were: 1) Poiseuille flow, 2) Couette flow, and 3) open boundary flow.

\begin{figure}
	\centering
	  \includegraphics[width=0.6\textwidth]{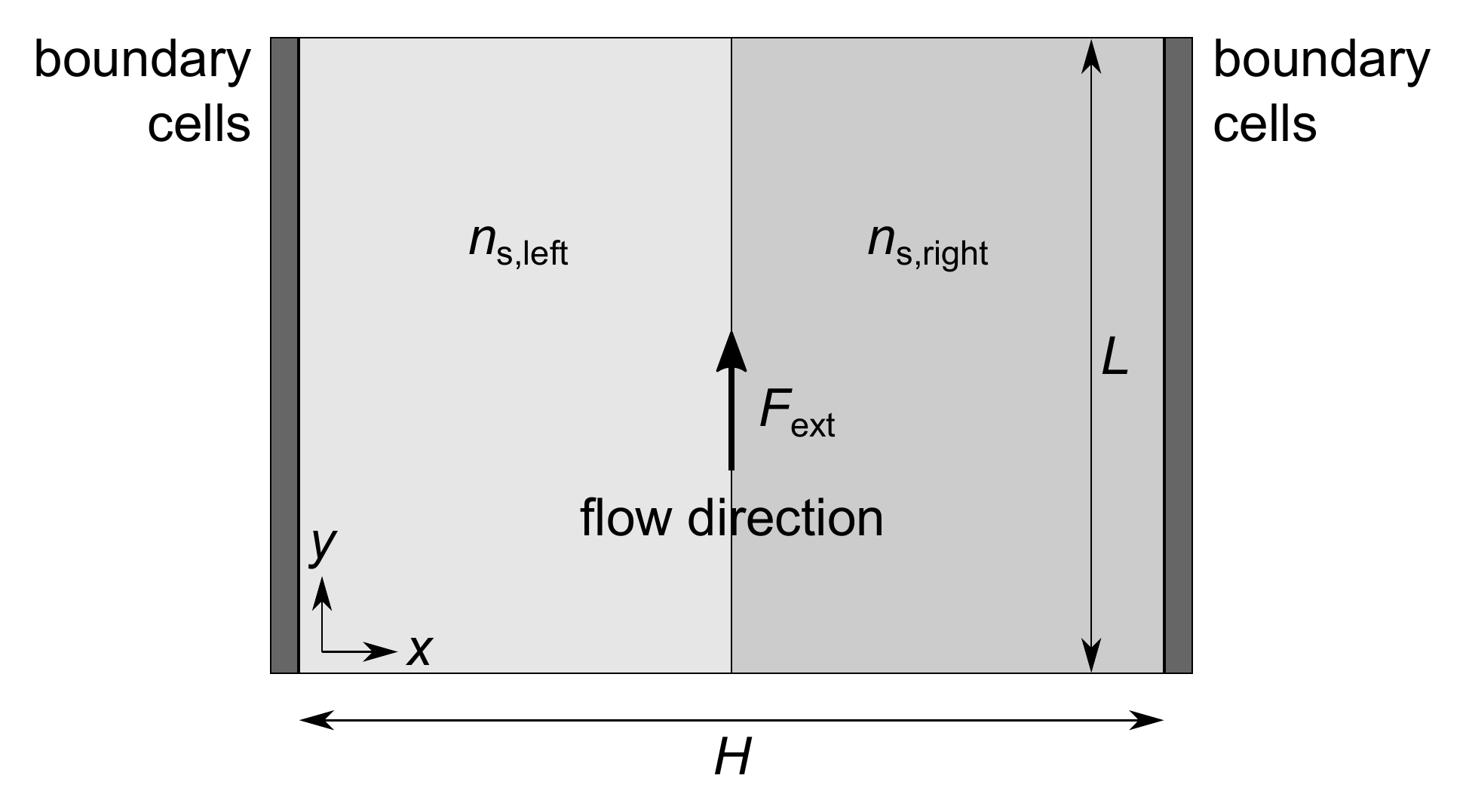}
	\caption{Schematic simulation setup for Darcy-Brinkman-type flow phenomena. The dimensions of the channel along the $x$- and $y$-directions are $H$ and $L$, respectively. The left ($x=[0,H/2]$, light gray) and the right ($x=]H/2,H]$, medium gray) half of the channel are filled with porous media defined by $n_{\mathrm{s},\mathrm{left}}$ and $n_{\mathrm{s},\mathrm{right}}$, respectively. The channel is bounded by boundary cells (dark gray) in the $x$-direction and has periodic boundaries in the $y$-direction. The flow is driven in $+y$-direction, e.g.\ by a body force $F_\mathrm{ext}$. }
	\label{fig:DarcyBrinkman}
\end{figure}

The overall simulation scenario is schematically shown in \autoref{fig:DarcyBrinkman}. It consisted of two stratified layers of porous media for which the permeabilities were independently adjusted by $n_{\mathrm{s},\mathrm{left}}$ and $n_{\mathrm{s},\mathrm{right}}$. The simulation domain had the dimensions $H$ and $L$ along the $x$- and $y$-directions, respectively. The flow was driven along the $+y$-direction either by applying constant velocities to the boundary cells or a body force $F_\mathrm{ext}$ to all lattice cells. At steady state, the velocity profile $u(x)$ was determined. 

\paragraph{1) Poiseuille flow} The Poiseuille flow was studied for a single-phase fluid in an homogeneous medium $\left(n_\mathrm{s}=n_{\mathrm{s},\mathrm{left}}=n_{\mathrm{s},\mathrm{right}}\right)$ for which $n_\mathrm{s}$ was varied in the range $n_\mathrm{s}=\{0.001,0.01,0.1,0.5,0.9\}$. No-slip boundary conditions were applied as bounce-back at the boundary cells. The dimensions and the body force were $H=50$\,lu, $L=50$\,lu, and $F_\mathrm{ext}=10^{-5}$\,(mu\,lu)/ts², respectively. The analytical solution of \autoref{eq:DarcyBrinkmanEq} for this case is
\begin{align}    
    \label{eq:DarcyBrinkmanPoiseuille}
    u = \frac{F_\mathrm{ext}k}{\nu}\left(1-\frac{\cosh\left[r(x-H/2) \right]}{\cosh(rH/2)}\right),
\end{align}
where $r=\sqrt{\nu\phi/k\nu_\mathrm{B}}=\sqrt{2n_\mathrm{s}/\nu}$ for $\phi=(1-n_\mathrm{s})$ and $\nu_\mathrm{B}=\nu$, and $k$ was determined using \autoref{eq:GS_perm}.

\begin{figure}
	\centering
	  \includegraphics[width=0.5\textwidth]{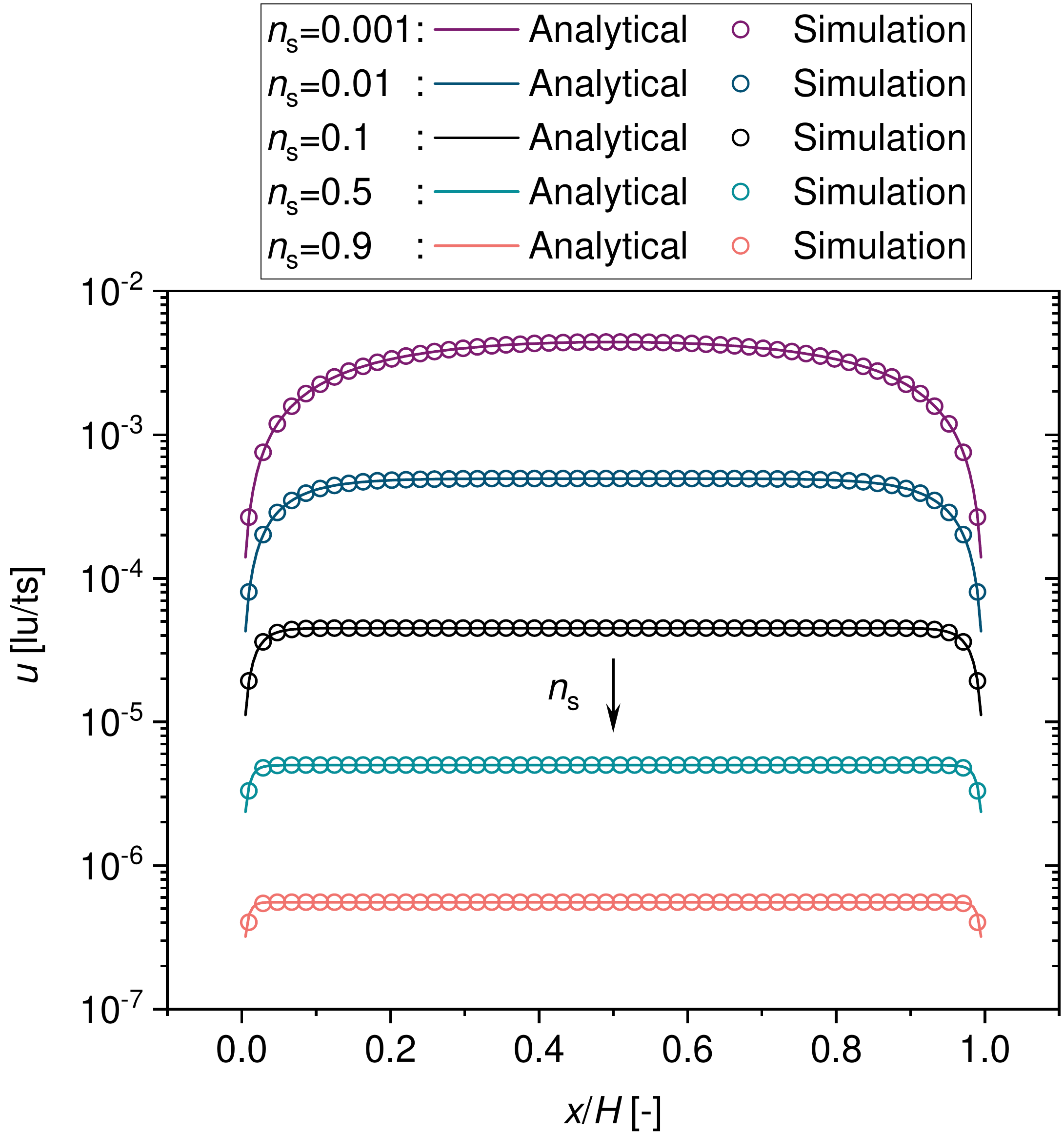}
	\caption{Velocity profiles of single-phase Darcy-Brinkman-type flow under Poiseuille conditions and at different bounce-back fractions $n_\mathrm{s}$. The dimensions of the channel are $H=50$\,lu and $L=50$\,lu. The flow is driven by the body force $F_\mathrm{ext}=10^{-5}$\,(mu\,lu)/ts². Simulation results (circles) are shown together with their analytical solutions (solid lines, cf.\ \autoref{eq:DarcyBrinkmanPoiseuille}). }
	\label{fig:DarcyBrinkmanPoiseuille}
\end{figure}

\autoref{fig:DarcyBrinkmanPoiseuille} shows the simulation results and analytical solutions (cf.\ \autoref{eq:DarcyBrinkmanPoiseuille}) of the velocity profiles. They are in excellent agreement over a wide range of $n_\mathrm{s}$, i.e.\ within Darcy $\left(\tilde{\sigma}=\mathrm{Da}^{-0.5}=(k/H^2)^{-0.5}\gtrsim100\right)$ and Brinkman $\left(100\gtrsim\tilde{\sigma}\gtrsim1\right)$ regime.

\paragraph{2) Couette flow} The Couette flow was also studied for a single-phase fluid in an homogeneous medium $\left(n_\mathrm{s}=n_{\mathrm{s},\mathrm{left}}=n_{\mathrm{s},\mathrm{right}}\right)$ for which $n_\mathrm{s}$ was varied in the range of $n_\mathrm{s}=\{0.001,0.01,0.1,0.5,0.9\}$ again. Velocity boundary conditions were applied to the boundary cells, i.e.\ $u(0)=0$\,lu/ts and $u(H)=U_0=0.01$\,lu/ts. The dimensions and the body force were $H=50$\,lu, $L=50$\,lu, and $F_\mathrm{ext}=0$\,(mu\,lu)/ts², respectively. The analytical solution of \autoref{eq:DarcyBrinkmanEq} for this case is given as
\begin{align}    
    \label{eq:DarcyBrinkmanCouette}
    u = U_0\frac{\sinh(rx)}{\sinh(rH)},
\end{align}
where again $r=\sqrt{2n_\mathrm{s}/\nu}$.

\begin{figure}
	\centering
	  \includegraphics[width=0.5\textwidth]{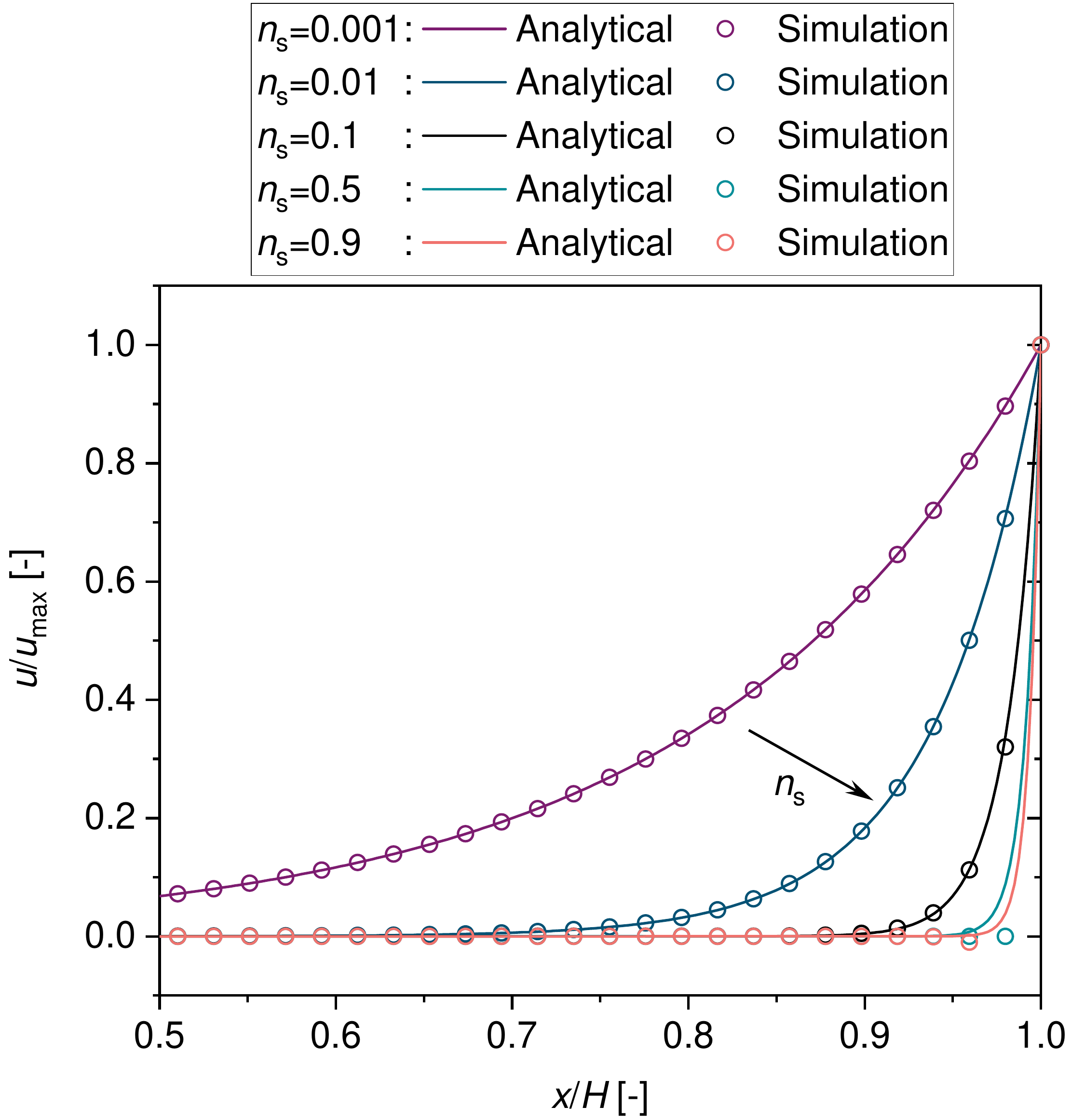}
	\caption{Velocity profiles of single-phase Darcy-Brinkman-type flow under Couette conditions and at different bounce-back fractions $n_\mathrm{s}$. The dimensions of the channel are $H=50$\,lu and $L=50$\,lu. The flow is driven by the wall velocity $U_0=u_{\mathrm{max}}=0.01$\,lu/ts. Simulation results (circles) are shown together with their analytical solutions (solid lines, cf.\ \autoref{eq:DarcyBrinkmanCouette}). }
	\label{fig:DarcyBrinkmanCouette}
\end{figure}

\autoref{fig:DarcyBrinkmanCouette} shows the simulation results and the analytical solutions (cf.\ \autoref{eq:DarcyBrinkmanCouette}) of the velocity profiles. They are in excellent agreement within the Brinkman regime, i.e.\ for $n_\mathrm{s}\leq0.1$. For larger values of $n_\mathrm{s}$, i.e.\ in the Darcy regime, deviations are observed close to the boundary cells $x/H\gtrsim0.95$. This was also reported for other GS models and explained with an $n_\mathrm{s}$-dependence of the no-slip conditions \cite{Chen2008,Zhu2013}.

\paragraph{3) Open boundary flow} The open boundary flow was studied for both a single-phase fluid and a two-phase fluid in an heterogeneous medium $\left(n_{\mathrm{s},\mathrm{left}}\neq n_{\mathrm{s},\mathrm{right}}\right)$. The influence of adhesion was neglected $\left(G_{\mathrm{ads},\sigma}=0.0\right)$. In the left half of the domain a constant value $n_{\mathrm{s},\mathrm{left}}=0.9$ was chosen, while $n_{\mathrm{s},\mathrm{right}}$ was varied in the range of $n_{\mathrm{s},\mathrm{right}}=\{0.001,0.01,0.1,0.5,0.8\}$. Instead of the boundary cells, periodic boundary conditions along the $x$-direction were applied. The dimensions and the body force were $H=100$\,lu, $L=50$\,lu, and $F_\mathrm{ext}=10^{-6}$\,(mu\,lu)/ts², respectively. In case of the two-phase flow, the system was initialized with fluid 1 in the left half $\left(x=[0,H/2]\right)$ and fluid 2 in the right half of the domain $\left(x=]H/2,H]\right)$. The piece-wise analytical solution of \autoref{eq:DarcyBrinkmanEq} for the open boundary flow is given as
\begin{subequations}
\label{eq:DarcyBrinkmanOpenBoundary}
    \begin{align}
      u_{\mathrm{l}} = U_{0,\mathrm{l}}\left[1-\frac{(1-p_\mathrm{l})\cosh(r_\mathrm{l}(x-\frac{H}{4}))}{\sinh(r_\mathrm{l}\frac{H}{4})\left[q_\mathrm{l}\coth(r_\mathrm{r}\frac{H}{4})+\coth(r_\mathrm{l}\frac{H}{4})\right]}\right], \forall x \in [0,\frac{1}{2}H], \label{eq:DarcyBrinkmanOpenBoundary1}
    \end{align}
    \begin{align}
      u_{\mathrm{r}} = U_{0,\mathrm{r}}\left[1+\frac{(p_\mathrm{r}-1)\cosh(r_\mathrm{r}(\frac{3H}{4}-x))}{\sinh(r_\mathrm{r}\frac{H}{4})\left[q_\mathrm{r}\coth(r_\mathrm{l}\frac{H}{4})+\coth(r_\mathrm{r}\frac{H}{4})\right]}\right], \forall x \in\;]\frac{1}{2}H,H]. \label{eq:DarcyBrinkmanOpenBoundary2}
    \end{align}
\end{subequations}
The corresponding parameters $U_0$, $p$, $q$, and $r$ are given in \autoref{tab:DarcyBrinkmanOpenBoundary}. The labels `l' and `r' denote the left $\left(x=[0,H/2]\right)$ and the right $\left(x=]H/2,H]\right)$ half of the simulation domain, respectively.
\begin{table}[h!]
	\centering
	\caption{Declaration of the parameters from \autoref{eq:DarcyBrinkmanOpenBoundary}.}
	\begin{tabular}{l||l}
		\toprule \toprule
        $U_{0,\mathrm{l}} = \frac{k_\mathrm{l}F_\mathrm{ext}}{\nu}$ & $U_{0,\mathrm{r}} = \frac{k_\mathrm{r}F_\mathrm{ext}}{\nu}$ \\
		$p_\mathrm{l} = \frac{k_\mathrm{r}}{k_\mathrm{l}}$ & $p_\mathrm{r} = \frac{k_\mathrm{l}}{k_\mathrm{r}}$ \\
		$q_\mathrm{l} = \sqrt{\frac{n_{\mathrm{s},\mathrm{left}}}{n_{\mathrm{s},\mathrm{right}}}}$ & $q_\mathrm{r} = \sqrt{\frac{n_{\mathrm{s},\mathrm{right}}}{n_{\mathrm{s},\mathrm{left}}}}$ \\
		$r_\mathrm{l} = \sqrt{\frac{2n_{\mathrm{s},\mathrm{left}}}{\nu}}$ & $r_\mathrm{r} = \sqrt{\frac{2n_{\mathrm{s},\mathrm{right}}}{\nu}}$ \\
		\bottomrule \bottomrule 
	\end{tabular}%
	\label{tab:DarcyBrinkmanOpenBoundary}%
\end{table}%

\begin{figure}
	\centering
	  \includegraphics[width=1.0\textwidth]{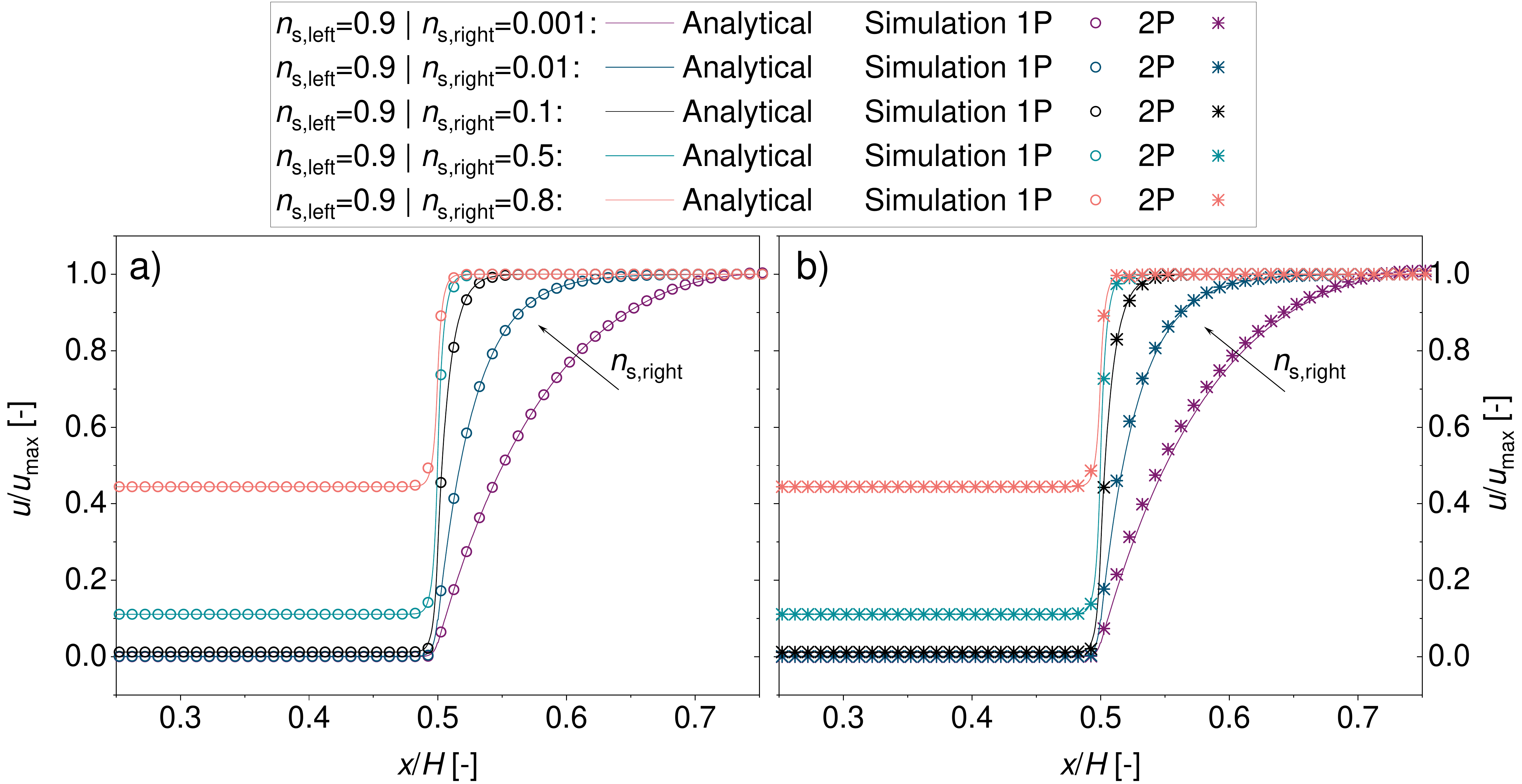}
	\caption{Velocity profiles for a) the single-phase and b) the two-phase Darcy-Brinkman-type flow under open boundary conditions. The velocity is normalized by the maximum velocity $u_{\mathrm{max}}$ for which the numerical values are given in \autoref{tab:DarcyBrinkmanGinzburg} in the Appendix. The bounce-back fraction $n_{\mathrm{s},\mathrm{left}}$ is kept constant at $n_{\mathrm{s},\mathrm{left}}=0.9$ , while $n_{\mathrm{s},\mathrm{right}}$ is varied in the range of $n_{\mathrm{s},\mathrm{right}}=\{0.001,0.01,0.1,0.5,0.8\}$. The dimensions of the channel are $H=100$\,lu and $L=50$\,lu. The flow is driven by the body force $F_\mathrm{ext}=10^{-6}$\,(mu\,lu)/ts². Simulation results for single-phase (circles) and two-phase flow (asterisks) are shown together with their analytical solutions (solid lines, cf.\ \autoref{eq:DarcyBrinkmanOpenBoundary}). }
	\label{fig:DarcyBrinkmanGinzburg}
\end{figure}

\autoref{fig:DarcyBrinkmanGinzburg} shows the simulation results for a) the single-phase flow and b) the two-phase flow as well as the corresponding analytical solutions (cf.\ \autoref{eq:DarcyBrinkmanCouette}) close to the interface ($x=[1/4,3/4]H$). The results of the single-phase flow are in excellent agreement over a wide range of $n_{\mathrm{s},\mathrm{right}}$, i.e.\ within Darcy and Brinkman regime. The results of the two-phase flow are also in good agreement with the analytical solutions. However, they are slightly overestimated for $n_{\mathrm{s},\mathrm{right}}<0.1$ due to slip at the interface. Moreover, the results indicate that the HMCSC inherently overcomes an issue of the GS-WBS, which was reported by Zhu and Ma \cite{Zhu2013} and Ginzburg \cite{Ginzburg2016}. They found that the GS-WBS in the formulation of \cite{Zhu2013} revealed a non-physical velocity discontinuity at the interface for cases where $n_{\mathrm{s},\mathrm{left}}\approx n_{\mathrm{s},\mathrm{right}}\approx1.0$. This was not observed here (cf. simulation with $n_{\mathrm{s},\mathrm{left}}=0.9$ and $n_{\mathrm{s},\mathrm{right}}=0.8$) and is assumed to be due to using the Shan-Chen forcing scheme in the present study (cf.\ \autoref{sec:Appendix_GuoShan}).

\subsection{Bubble Test} \label{sec:BubbleTest}
The interfacial tension $\gamma$ is an inherent thermodynamic property that depends on the molecular interaction of a set of immiscible fluids as well as on the thermodynamic state of the system. Therefore, unlike in other GS models \cite{McDonald2016,Pereira2016}, it should not depend on the homogenization of a nanoporous medium. This is especially important, when the medium is heterogeneous, i.e.\ $n_\mathrm{s}$ is space-dependent. Thus, for each homogenized multi-phase LB model to be physically consistent, it has to be ensured that setting the model parameter $G_{\mathrm{inter},12}$ to a constant value for the whole simulation domain, does not lead to spatially varying interfacial tensions.

\begin{figure}
	\centering
	  \includegraphics[width=1.0\textwidth]{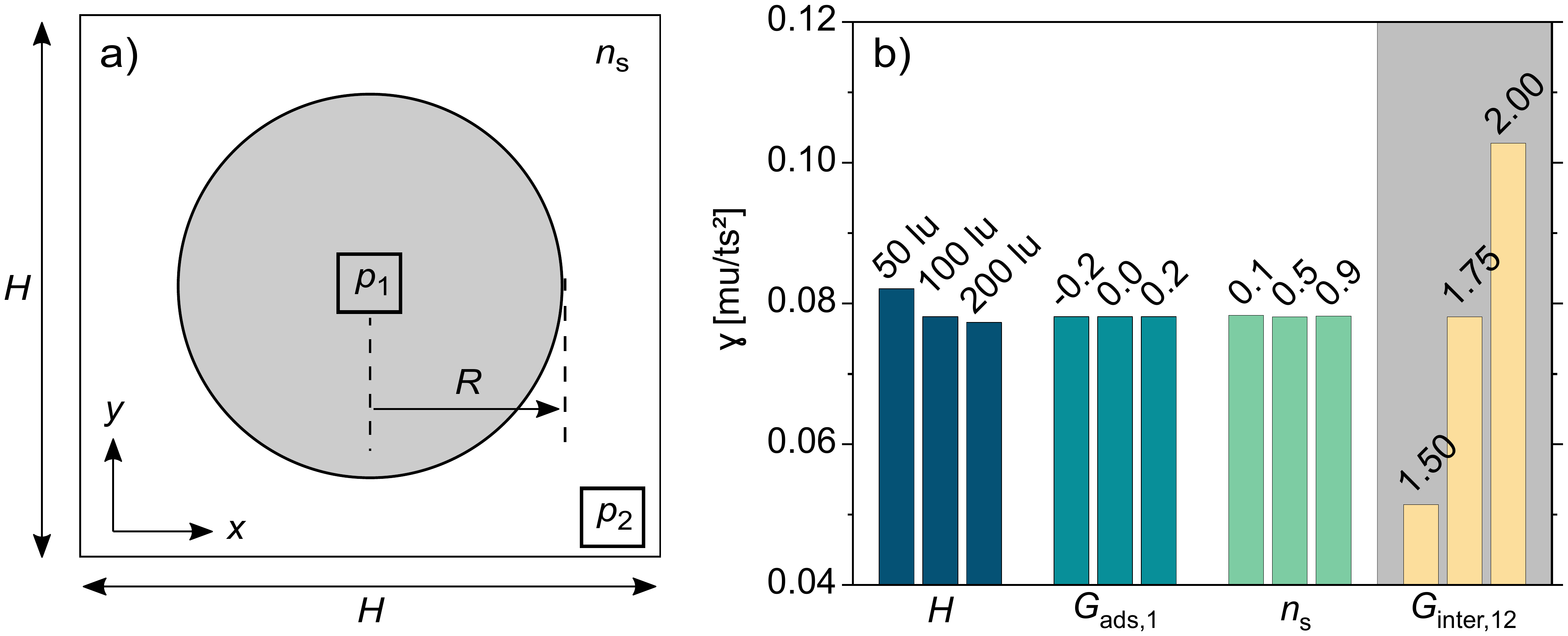}
	\caption{Bubble test. a) Simulation setup in which a bubble of fluid 1 (gray) is submersed in fluid 2 (white). The domain size is $H$ along the $x$- and $y$-directions. The radius $R$ and the pressures of both fluids, i.e.\ $p_1$ and $p_2$, were determined to calculate the interfacial tension $\gamma$. b) Simulation results of $\gamma$ as a function of different influencing factors, for which the specific values are given in the plot. }
	\label{fig:Laplace}
\end{figure}

This was verified using bubble tests. The simulation setup is shown in \autoref{fig:Laplace}a). The simulation domain consisted of an homogenized porous medium where all lattice cells had the identical bounce-back fraction $n_\mathrm{s}$. The system was fully periodic. The dimensions of the simulation domain along the $x$- and $y$-direction were $H$. The system was initially filled with a bubble consisting of fluid 1 which was submersed in fluid 2. Both components had equal masses. The pressure difference between the center of the bubble and its surroundings, i.e.\ $\Delta p=p_\mathrm{1}-p_\mathrm{2}$, as well as the bubble radius $R$ were determined from the simulations. Therefrom, $\gamma$ follows Laplace's law $\gamma = \Delta p R$. The influence of the parameters $H$, $G_{\mathrm{ads},1}$, $n_\mathrm{s}$, and $G_{\mathrm{inter},12}$ on $\gamma$ was studied. For the reference simulation case, the parameters were $H=100$\,lu, $G_{\mathrm{ads},1}=0.0$, $n_\mathrm{s}=0.5$, and $G_{\mathrm{inter},12}=1.75$. Unless specified otherwise, the subsequent simulations used those default parameters.

\autoref{fig:Laplace}b) shows the results of $\gamma$ for different values of $H$, $G_{\mathrm{ads},1}$, $n_\mathrm{s}$, and $G_{\mathrm{inter},12}$. It was observed that $\gamma$ does not depend on $G_{\mathrm{ads},1}$ and $n_\mathrm{s}$. This was also confirmed for bubble spreading in structurally resolved porous media and is shown in \autoref{fig:nonHomBubbleTest} in the Supporting Information. Thus, the interfacial tension was only affected by $H$ and $G_{\mathrm{inter},12}$. For increasing $H$, $\gamma$ decreased slightly and converged towards the asymptotic value $\gamma=0.077\,\mathrm{mu/ts}^2$. This was related to the imprecise determination of the bubble radius $R$ for small systems, but worked well for $H\geq100$\,lu. In contrast, the influence of the fluid-fluid interaction $G_{\mathrm{inter},12}$ on $\gamma$ was large which was to be expected. $\gamma$ increased with increasing $G_{\mathrm{inter},12}$. The results were in perfect agreement with the corresponding values reported by Huang~\textit{et al.} \cite{Huang2007}. These were $\gamma=\{0.051,0.078,0.104\}\,\mathrm{mu/ts}^2$ for $G_{\mathrm{inter},12}=\{1.50,1.75,2.00\}$, respectively. They were determined using the original MCSC, which corresponds to using the HMCSC for $n_\mathrm{s}=0$. 

\subsection{Washburn Simulations} \label{sec:Washburn}
The Washburn equation \cite{Washburn1921}
\begin{align}    
    \label{eq:Washburn}
    x_{\mathrm{inter}} = \sqrt{\frac{\gamma \cos(\theta)}{2\eta}} \sqrt{R_{\mathrm{eff}} t}=A\sqrt{Bt},
\end{align}
describes imbibition into homogeneous and isotropic porous media as 1D flow through a bundle of cylindrical tubes with effective radius $R_{\mathrm{eff}}$. The position of the moving interface is denoted by $x_{\mathrm{inter}}$. The equation is only valid if gravity is negligible \cite{Das2013,Li2015}.

The flow is driven by the interplay of capillary forces and viscous forces, where the former depend on the interfacial tension $\gamma$ and the contact angle $\theta$, and the latter are determined by the dynamic viscosity of the fluid $\eta=\rho\nu$. The parameters $A$ and $B$ in \autoref{eq:Washburn} are fit parameters that are used in the following to fit the simulation data to the Washburn equation.

The simulation scenario consisted of a homogenized porous medium in which all lattice cells had the identical bounce-back fractions $n_\mathrm{s}$. The system dimensions along the $x$- and $y$-direction were $H=500$\,lu and $L=5$\,lu, respectively. The system was initially filled with fluid 2. The densities of fluid 1 and fluid 2 were prescribed at the inlet $\left(\rho_1(x=0)\right)$ and outlet $\left(\rho_2(x=H)\right)$, respectively. Periodic boundary conditions were applied along the $y$-direction. Fluid 1 penetrated into the simulation domain in $+y$-direction and displaced fluid 2. The influence of the solid-fluid interaction and bounce-back fraction was studied. Correspondingly, the contact angle $\theta$ was varied by changing $G_{\mathrm{ads}}$ in the range of $G_{\mathrm{ads}}=-G_{\mathrm{ads},1}=[0.05,0.40]$, and the permeability or effective radius $R_{\mathrm{eff}}$ was varied by changing $n_\mathrm{s}$ in the range of $n_\mathrm{s}=\{0.1,0.5,0.9\}$.

\begin{figure}
	\centering
	  \includegraphics[width=1.0\textwidth]{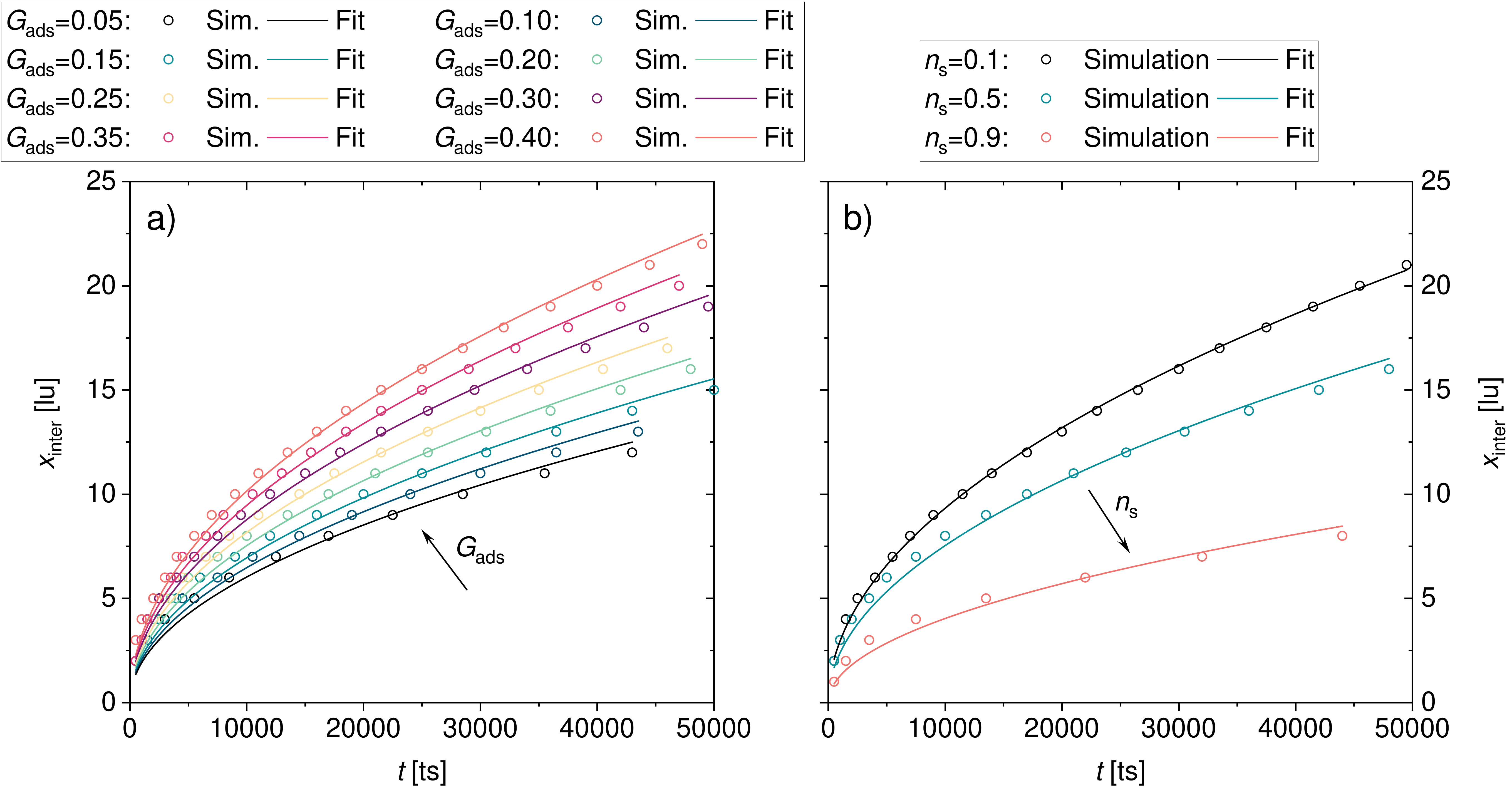}
	\caption{Washburn simulations. Interface position $x_{\mathrm{inter}}$ as a function of time. The influence of a) the solid-fluid interaction $G_{\mathrm{ads}}$ and b) the bounce-back fraction $n_\mathrm{s}$ were studied. Simulation results (circles) are compared with fits to the analytical solutions of the Washburn equation (solid lines, cf.\ \autoref{eq:Washburn}) for which the parameters $A$ and $B$ are given in \autoref{tab:Washburn}.}
	\label{fig:Washburn}
\end{figure}

\autoref{fig:Washburn} shows the simulation results for different values of $G_{\mathrm{ads}}$ and $n_\mathrm{s}$. Results are given for $t\leq5\cdot 10^4$\,ts. This corresponds to the Washburn regime $\left(x_{\mathrm{inter}}\sim\sqrt{t}\right)$ where viscous effects dominate the flow \cite{Das2012,Das2013,Li2015}. The data was used for fitting the Washburn equation \autoref{eq:Washburn}. 

For the simulations shown in \autoref{fig:Washburn}a), $n_\mathrm{s}=0.5$ and only $G_{\mathrm{ads}}$ was varied, while for the simulations in \autoref{fig:Washburn}b), $G_{\mathrm{ads}}=0.2$ and only $n_\mathrm{s}$ was varied. Correspondingly, for all fits in \autoref{fig:Washburn}a), parameter $B$ was the same and only $A$ was freely adapted. Vice versa for all fits in \autoref{fig:Washburn}b), $A$ was fixed and only $B$ was fitted to the simulations. Overall, the fits and the data are in good accordance. This is also confirmed by the large coefficients of determination, i.e.\ $R_A^2,R_B^2>0.97$.

The values for $A$ and $B$ as well as $\theta$ and $R_{\mathrm{eff}}$ are given in \autoref{tab:Washburn}. In addition, $\theta$ determined from the Washburn simulations was compared with the correlation for the contact angle of Huang~\textit{et al.}\ \cite{Huang2007}
\begin{align}
    \label{eq:contactAngle}
   \cos(\theta_{\mathrm{H}}) = \frac{4G_{\mathrm{ads},2}}{G_{\mathrm{inter},12}(\rho_1-\rho_{\mathrm{dis},2})},
\end{align}
which was derived from sessile droplet simulations using the original MCSC. Interestingly, a good accordance of $\theta$ and $\theta_{\mathrm{H}}$ was observed, especially for $\theta\in[0,60]\degree$. For larger values of $\theta$, the correlation slightly overestimated the simulation results. This was to some extent also observed in the paper of Huang~\textit{et al.}\ \cite{Huang2007} and is amplified by the increasing uncertainty of the fitting for $\theta>60\degree$ (cf. $R_A^2$ in \autoref{tab:Washburn}). 

\begin{table}[h!]
	\centering
	\caption{Fit parameters $A$ and $B$ determined from the simulations (cf.\ \autoref{eq:Washburn}). The uncertainty of the fit is estimated by the corresponding coefficients of determination, i.e.\ $R_A^2$ and $R_B^2$, respectively. The data of $\theta$ and $R_{\mathrm{eff}}$ determined from the simulations using the Washburn equation \autoref{eq:Washburn} are given. They are compared to $\theta_{\mathrm{H}}$ using the correlation by Huang~\textit{et al.} \cite{Huang2007} (cf.\ \autoref{eq:contactAngle}). }
	\begin{tabular}{l c c r r||l c c r}
		\toprule \toprule
        $G_{\mathrm{ads}}$ & $A$ $\left[\frac{\sqrt{\mathrm{lu}}}{\sqrt{\mathrm{ts}}}\right]$ & $R_A^2$ & $\theta$ [\degree] & $\theta_{\mathrm{H}}$ [\degree] & $n_\mathrm{s}$ & $B$ [lu] & $R_B^2$ & $R_{\mathrm{eff}}$ [lu] \\
        \midrule
		0.05 & 0.275 & 0.979 & 70.8 & 83.3 & 0.1 & 0.0737 & 0.999 & 0.0737 \\
		0.10 & 0.295 & 0.984 & 67.7 & 76.5 & 0.5 & 0.0481 & 0.990 & 0.0481 \\
		0.15 & 0.317 & 0.988 & 64.1 & 69.5 & 0.9 & 0.0138 & 0.972 & 0.0138 \\
		0.20 & 0.344 & 0.990 & 59.2 & 62.2 &&& \\
		0.25 & 0.373 & 0.992 & 52.9 & 54.3 &&& \\
		0.30 & 0.401 & 0.995 & 45.9 & 45.6 &&& \\
		0.35 & 0.432 & 0.996 & 36.0 & 35.3 &&& \\
		0.40 & 0.463 & 0.995 & 21.4 & 21.1 &&& \\
		\bottomrule \bottomrule 
	\end{tabular}%
	\label{tab:Washburn}%
\end{table}%

\subsection{Porous Obstacle Flow} \label{sec:PorousObstacle}
The last test concerned the simultaneous occurrence of free flow and Darcy-Brinkman flow. It followed an example of Spaid and Phelan \cite{Spaid1998} which is also typically used as advanced benchmark in the literature \cite{Yoshida2014,Silva2015,Pereira2016}.

\begin{figure}
	\centering
	  \includegraphics[width=0.5\textwidth]{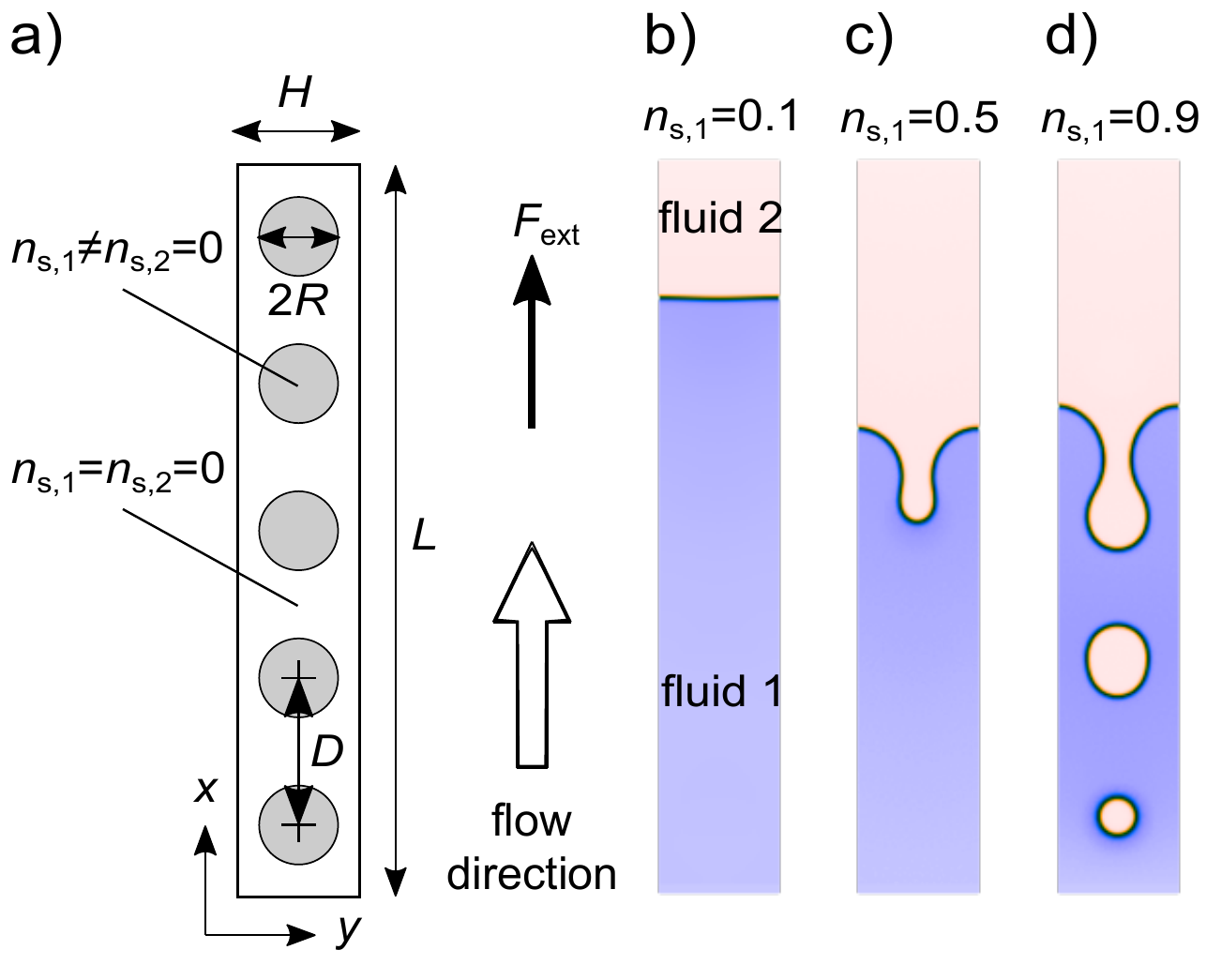}
	\caption{Two-phase flow through a series of circular porous obstacles. a) Schematic simulation setup. The dimensions along the $x$- and $y$-direction are $L$ and $H$, respectively. The system contains five circular porous obstacles (gray shaded area) with radius $R$ that are equidistantly distributed by the distance $D$. Within the obstacles $n_{\mathrm{s},1}$ is varied in the range of $n_{\mathrm{s},1}=\{0.1,0.5,0.9\}$. A body force $F_\mathrm{ext}$ is applied along the $+x$-direction. b)$-$d) Exemplary simulation results at different values of $n_{\mathrm{s},1}$. The snapshots show the distribution of fluid 1 (blue) and fluid 2 (red) at the time step $t=50{,}000$\,ts each. }
	\label{fig:PorousObstacleScheme}
\end{figure}

The simulation scenario is shown in \autoref{fig:PorousObstacleScheme}a). It consisted of a channel with five circular porous obstacles with diameter $2R=33$\,lu from which the center-to-center distance was $D=60$\,lu. The dimensions of the system along the $x$- and $y$-direction were $L=300$\,lu and $H=50$\,lu, respectively. Periodic boundary conditions were applied along the $y$-direction. The system was initially filled with fluid 2. The densities of fluid 1 and fluid 2 were prescribed at the inlet $\left(\rho_1(x=0)\right)$ and  outlet $\left(\rho_2(x=L)\right)$, respectively. A body force $F_\mathrm{ext}=5\cdot 10^{-5}$\,(mu\,lu)/ts² in $+x$-direction was applied to both fluids, leading to fluid 1 penetrating into the simulation domain.

To ensure consistency with the simulations of Spaid and Phelan \cite{Spaid1998}, viscous effects of fluid 2 within the obstacles were neglected by setting $n_{\mathrm{s},2}=0$ in all lattice cells of the domain. In contrast, $n_{\mathrm{s},1}$ was set to $n_{\mathrm{s},1}=\{0.1,0.5,0.9\}$ within the obstacles to account for the Darcy-Brinkman-type behavior of fluid 1 and to $n_{\mathrm{s},1}=0$ for all other lattice cells. Thus, large viscosity ratios between the two fluids in the porous obstacles could be mimicked without running into numerical instabilities. 

This scenario can be mathematically described by Darcy's law
\begin{align}
    \label{eq:PorousObstacle}
    \frac{\mathrm{d}x_{\mathrm{inter}}}{\mathrm{d}t} = \frac{k_{\mathrm{unsat}}}{\phi \eta} F_\mathrm{ext},
\end{align}
where $x_{\mathrm{inter}}$ is the interface position, $k_{\mathrm{unsat}}$ is the permeability of the unsaturated system, $\phi$ is the effective porosity of the obstacles, $\eta$ is the dynamic viscosity of fluid 1 and $F_\mathrm{ext}$ is the body force. The analytical solution of \autoref{eq:PorousObstacle} is
\begin{align}
    \label{eq:PorousObstacleSolution}
   x_{\mathrm{inter}} = \frac{k_{\mathrm{unsat}}}{\phi \eta} F_\mathrm{ext}t,
\end{align}
which was used to determine $k_{\mathrm{unsat}}$ from the simulations.

Representative snapshots of the simulations at $t=50{,}000$\,ts are shown in \autoref{fig:PorousObstacleScheme}b)$-$d) for different values of $n_{\mathrm{s},1}$. They reveal that increasing $n_{\mathrm{s},1}$ retards the filling of the channel. For $n_{\mathrm{s},1}=0.9$ even a distinct bubble or droplet formation within the porous obstacles was observed. These results agree qualitatively with results from the literature \cite{Spaid1998,Pereira2016}.

\begin{figure}
	\centering
	  \includegraphics[width=0.6\textwidth]{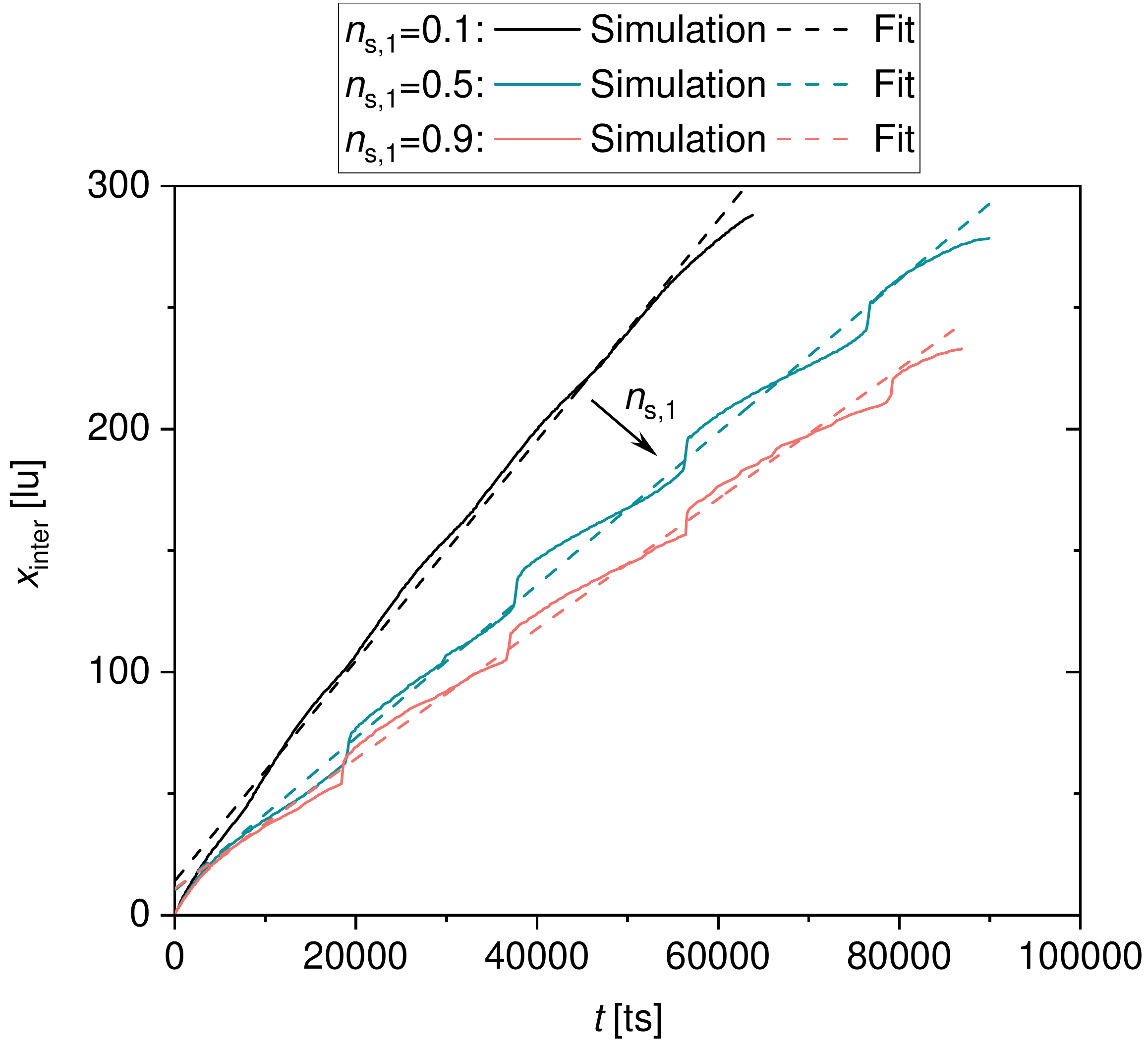}
	\caption{Interface position $x_{\mathrm{inter}}$ for a two-phase flow through a series of circular porous obstacles. Results are shown as a function of time $t$ and $n_{\mathrm{s},1}$. Simulation results (solid lines) are compared with the fits to the analytical solutions (dashed lines, cf.\ \autoref{eq:PorousObstacleSolution}) for which $k_{\mathrm{unsat}}$ is given in \autoref{tab:PorousObstacle}.}
	\label{fig:PorousObstacleResults}
\end{figure}

Complementary, \autoref{fig:PorousObstacleResults} shows quantitative simulation results and the corresponding fits to \autoref{eq:PorousObstacleSolution} for different values of $n_{\mathrm{s},1}$. The position of the interface was determined as $x_{\mathrm{inter}} = L/2 \left(S_1(y=0)+ S_1(y=H/2)\right)$, where the proportions of cells in the boundary and the center column of the simulation domain which were filled with fluid 1 were denoted $S_1(y=0)$ and $S_1(y=H/2)$, respectively. The oscillations around the linear fits correspond to alternating deceleration and acceleration of the fluid front which depends on the varying width of the free-flow channel. The unsaturated permeabilities $k_{\mathrm{unsat}}$ were determined fitting \autoref{eq:PorousObstacleSolution} to the simulation data. The corresponding permeability of a saturated medium $k_{\mathrm{sat}}$ was determined following the lubrication theory of Phelan and Wise \cite{Phelan1996}. 

Both $k_{\mathrm{unsat}}$ and $k_{\mathrm{sat}}$ are given in \autoref{tab:PorousObstacle}. The data shows that $k_{\mathrm{unsat}}<k_{\mathrm{sat}}$ and its ratio $k_{\mathrm{unsat}}/k_{\mathrm{sat}}$ decreases for increasing $n_{\mathrm{s},1}$. These results are in qualitative agreement with simulative \cite{Spaid1998,Yoshida2014} and experimental \cite{Parnas1995} observations.

\begin{table}[!ht]
	\centering
	\caption{Comparison of unsaturated and saturated permeabilities, i.e.\ $k_{\mathrm{unsat}}$ and $k_{\mathrm{sat}}$, respectively. }
	\begin{tabular}{l c c c}
		\toprule \toprule
        $n_{\mathrm{s},1}$ & $k_{\mathrm{unsat}}$ [lu²] & $k_{\mathrm{sat}}$ [lu²] & $k_{\mathrm{unsat}}/k_{\mathrm{sat}}$ \\
        \midrule
		0.1 & 10.71 & 35.86 & 0.299\\
		0.5 &  7.42 & 29.38 & 0.253\\
		0.9 &  6.31 & 27.59 & 0.229\\
		\bottomrule \bottomrule 
	\end{tabular}%
	\label{tab:PorousObstacle}%
\end{table}%

\subsection{Discussion} \label{sec:Discussion} 
In this section, the HMCSC was successfully validated using different single- and two-phase flow benchmark scenarios. Altogether, they demonstrate the broad applicability of the method presented here. Moreover, and just as important, it was shown that the HMCSC inherits all positive attributes from the original MCSC, the GS-WBS, and its multi-phase extension by Pereira \cite{Pereira2016}, while overcoming some deficiencies of these methods.
The features of the HMCSC are briefly summarized as follows: 
\begin{enumerate}
\item As was shown by the bubble tests, the interfacial tension $\gamma$ is independent of $n_\mathrm{s}$ and $G_{\mathrm{ads}}$. Thus, the interfacial tension is a property of the fluids only and is not affected by the homogenization. 
\item As was shown by the Washburn simulations, also the contact angle $\theta$ is independent of $n_\mathrm{s}$. It is a property of the solid-fluid material combination only and is also not affected by the homogenization.
\item As was shown by the open boundary Darcy-Brinkman flow simulations, the HMCSC inherently ensures velocity continuity at interfaces of porous media with different $n_\mathrm{s}$. No additional smoothing procedure as suggested by Yehya~\textit{et al.} \cite{Yehya2015} is required.
\item The HMCSC is fully consistent with the original MCSC. Thus, the values for $G_{\mathrm{ads}}$  and $G_{\mathrm{inter}}$ can be chosen identical to the values that are used for the original MCSC. The same parametrization approach following the paper of Huang~\textit{et al.}\ \cite{Huang2007} as well as \autoref{eq:contactAngle} can be applied to study the identical physical situations. No further parametrization is required.
\item The HMCSC is especially suitable for studying multi-phase flow in heterogeneous porous media where both the wetting properties and the permeability vary in space and time, while the physical properties of the fluid mixture are unaffected.
\item The HMCSC is accurate, intuitive, easy to implement, and no stability issues have been observed for the parameter ranges studied here.
\end{enumerate}

Moreover, in contrast to most other GS models \cite{Thorne2004,Chen2008,Zhu2013,Zhu2018} including the model proposed by Pereira \cite{Pereira2016} where forces were included using the Guo forcing scheme \cite{Guo2002a}, the HMCSC uses the common Shan-Chen forcing scheme. Although, this leads to a $\tau$-dependence of the viscosity \cite{Yu2010,Silva2015}, it involves substantial advantages over the aforementioned models. The corresponding differences and their potential effects are briefly discussed in \autoref{sec:Appendix_GuoShan}.

\section{Application and Results} \label{sec:Results} 
Practical applications of the HMCSC are most scenarios in which multi-phase fluid flow occurs in multi-scale porous media. In the context of hydrology, geoscience and petroleum engineering, the HMCSC can be especially interesting to study transport in pores, vugs and microfractures simultaneously, while also considering local changes of the permeability due to geochemical or biological processes. Thus, the HMCSC might be helpful to predict and gain insight into groundwater hydrology, geologic carbon storage and sequestration, and the recovery of oil and gas from different multi-scale porous rocks, such as sandstones, carbonates, and shale \cite{Krevor2012,Ghezzehei2012,Mehmani2015,Soulaine2019,Mehmani2020,Hassannayebi2021}.

However, as the research focus of our group is on energy storage materials, here the electrolyte filling of lithium-ion batteries was studied exemplarily. The pore sizes in such microstructures typically range from nano- to micrometers and its filling is not yet fully understood. Thus, it is of recent research interest with the objective to optimize the corresponding manufacturing process as well as the battery performance and lifetime \cite{Wood2015,Weydanz2018}.

A realistic 3D reconstruction of a lithium-ion battery cathode \cite{Westhoff2018} with a porosity of $\phi_{\mathrm{A}}=40\,\%$ and neutral wetting conditions was used as a geometrical basis for all simulations. The filling of three variants of this structure was studied: 1) The pure electrode structure, 2) the electrode structure infiltrated with $V_{\mathrm{B}}=21\,\%$ volume fraction of a nanoporous binder, and 3) the electrode structure attached to a nanoporous and fully homogenized separator for which the generation is described in \autoref{sec:Data}. 

The contact angles of the active material, binder, and separator were $\theta_{\mathrm{A}}=90\degree$, $\theta_{\mathrm{B}}=60\degree$, and $\theta_{\mathrm{S}}=90\degree$, respectively. They were converted to the model parameter $G_{\mathrm{ads},2}=-G_{\mathrm{ads},1}$ using \autoref{eq:contactAngle}. The bounce-back fractions in the homogenized porous media were $n_\mathrm{s}=n_{\mathrm{s},1}=n_{\mathrm{s},2}=0.5$ for the binder, while in the separator $n_\mathrm{s}$ was space-dependent and identical to the normalized grayscale value of each voxel of the separator image data.

\begin{figure}
	\centering
	  \includegraphics[width=1.0\textwidth]{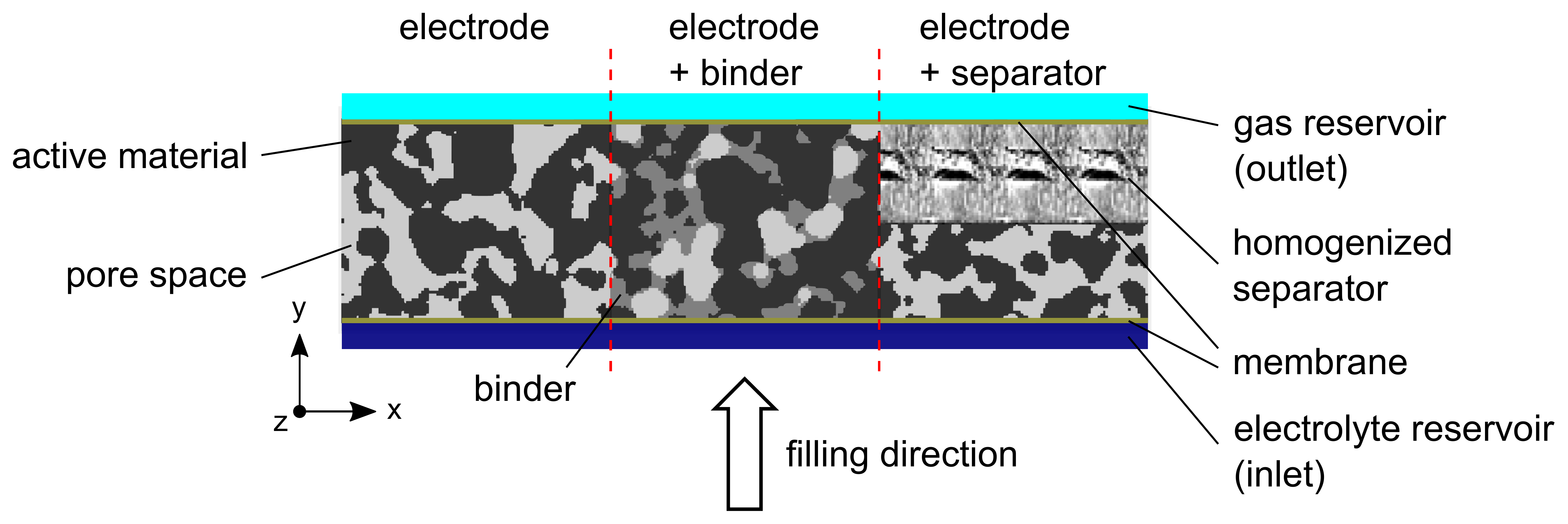}
	\caption{Simulation setup schematically shown for the three different electrode structures. The pure electrode (left) consists of active material (black) and mesoscopic pore space (light gray) only. For the variant with binder (middle), the binder (darker gray) is infiltrated into this pore space. For the variant with separator (right), the separator (grayscale reconstruction) replaces the upper half of the electrode. The reservoirs in which the densities of electrolyte and gas are prescribed are marked in blue and cyan, respectively. The semi-permeable membranes adjacent to the reservoirs are depicted in yellow. }
	\label{fig:SystemSetup}
\end{figure}

\autoref{fig:SystemSetup} shows the schematic simulation setup where also the three different variants of microstructures are indicated. It is similar to the setup that was used in a recent study of our group \cite{Lautenschlaeger2022} in which the electrolyte filling process was studied in very detail. Initially, the pore space was filled with gas only. Periodic boundary conditions were applied along the $x$- and $z$-direction. Along the $y$-direction an electrolyte reservoir and a gas reservoir were added at the inlet and outlet, respectively. The reservoirs had a thickness of four layers each in which the density of both fluids was prescribed. The initial electrolyte density at the inlet was $\rho_1$ and was incrementally increased during the simulation run. The gas density at the outlet was constant, i.e.\ $\rho_2$. Thereby, a pressure difference between the two fluids was applied to drive the electrolyte imbibition (cf. \autoref{eq:CapPressure}). Between the reservoirs and the microstructures semi-permeable membranes were placed to prevent an unwanted fluid breakthrough. The inlet membrane was permeable for the electrolyte only. The outlet membrane was permeable for the gas only. This approach is in accordance with imbibition experiments and simulations that are typically used to analyze porous media in the context of geoscience or energy storage materials \cite{Gostick2008,Karpyn2009,Pini2012,Krevor2012,Zhao2015,Danner2016,Tavangarrad2019,Zhu2021,Lautenschlaeger2022}. 

The model parameters for the simulations are given in \autoref{tab:Parameters}, where fluid 1 and fluid 2 correspond to electrolyte and gas, respectively.

From each simulation, pressure-saturation curves were determined. They are a characteristic property of porous media and relate the pressure difference $\Delta p$ needed for the imbibition to the amount of electrolyte in the pore space, i.e.\ the saturation $S_1$. The pressure difference $\Delta p$ was determined as
\begin{align}    
    \label{eq:CapPressure}
    \Delta p = \left<p\right>_{\text{inlet}} - \left<p\right>_{\text{outlet}},
\end{align}
where $p$ was evaluated using \autoref{eq:SC_pressure}, and $\left<p\right>$ denotes the average pressure in the inlet and outlet reservoirs. The electrolyte saturation $S_1$ was determined as
\begin{align}    
    \label{eq:Saturation}
    S_1 = \frac{N_{\mathrm{pore}}(\rho_1\geq0.5)+(1-n_\mathrm{s})N_{\mathrm{hom}}(\rho_1\geq0.5)}{N_{\text{pore}}+(1-n_\mathrm{s})N_{\mathrm{hom}}},
\end{align}
where the denominator and numerator correspond to the total pore space and the pore space in which $\rho_1\geq0.5$\,mu/lu$^3$, respectively. The number of pore lattice cells in the electrode structures and the homogenized nanoporous components are denoted by $N_{\mathrm{pore}}$ and $N_{\mathrm{hom}}$, respectively. The latter were multiplied by the effective fluid fraction $(1-n_\mathrm{s})$. Note, that for the calculation of the saturation only the lattice cells between the two membranes were considered.

A simulation run consisted of approximately 1,000,000 time steps. The pressure difference and the saturation were determined every 10,000 time steps during the production run. The simulations stopped when a further saturation was not possible and led to a steep increase of $\Delta p$. 

\begin{figure}
	\centering
	  \includegraphics[width=1.0\textwidth]{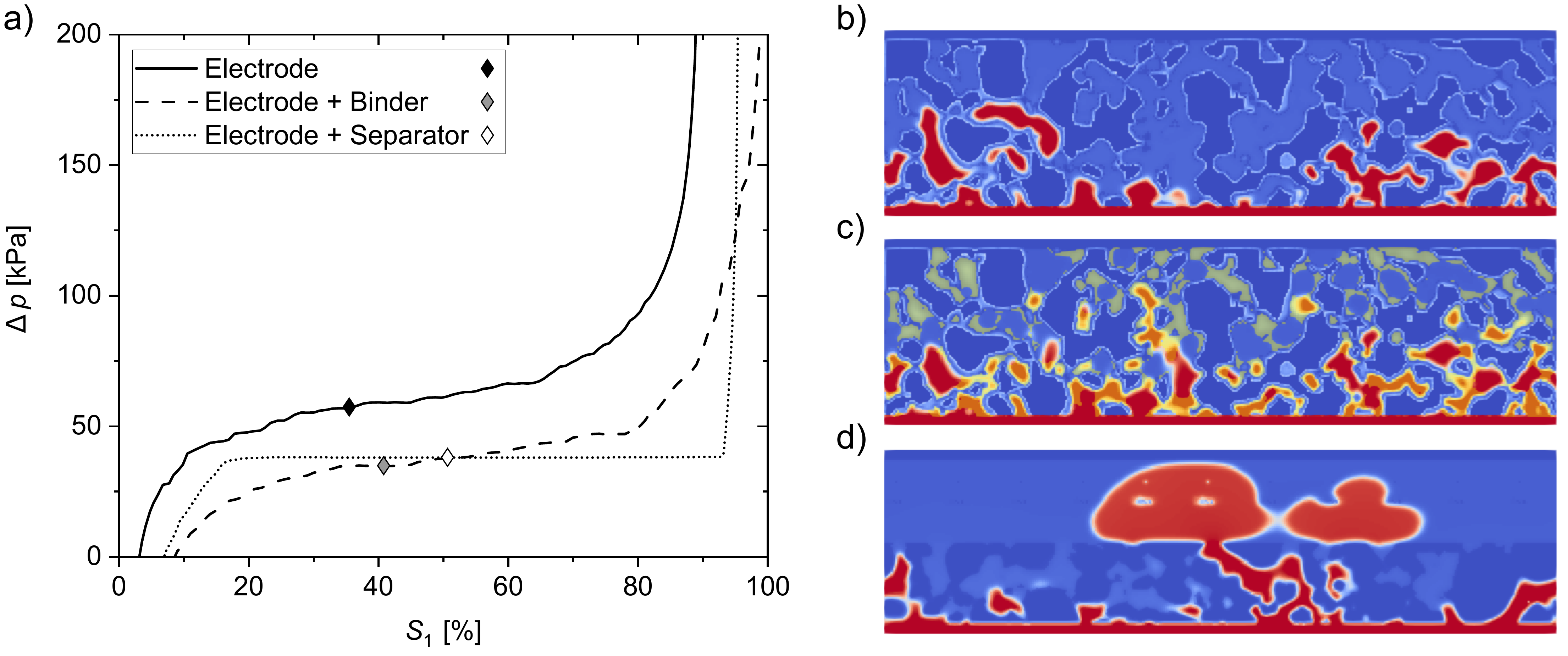}
	\caption{Electrolyte filling of electrode structures. a) Pressure-saturation behavior of the pure electrode (solid line), the electrode with binder (dashed line), and the electrode attached to the separator (dotted line). The hashes denote the state at time step $t=400{,}000$\,ts for which snapshots are shown on the right. b)$-$d) Snapshots of cross sections in the $xy$-plane at $t=400{,}000$\,ts for b) the pure electrode, c) the electrode with binder, and d) the electrode attached to the separator. The active component is depicted dark blue, the gas is depicted blue, the electrolyte is depicted red, and the binder is depicted yellow.  }
	\label{fig:pS}
\end{figure}

\autoref{fig:pS}a) shows the pressure-saturation curves for the three different microstructures. They follow a sigmoidal behavior which is typically observed in the literature and can be explained based on the Young–Laplace equation. For all cases, the final saturation $S_{\mathrm{1,final}}$ deviated from the theoretical optimum of 100\,\% which indicates residual gas being entrapped in the pore space \cite{Weydanz2018,Sauter2020}.

A remarkable influence of the homogenized microstructures was observed. Compared to the pure electrode, the pressure-saturation curve for the electrode with binder shows a similar qualitative behavior, but appears to be shifted by $\Delta p<0$ and $S_1>0$. Thus, by infiltrating a binder, the characteristics of the pore space were largely maintained. Only the enhanced wetting of the binder facilitated the electrolyte imbibition, led to a reduction of $\Delta p$ and increased $S_{\mathrm{1,final}}$ close to 100\,\%. The pressure-saturation curve for the electrode attached to the separator shows a stronger influence as also the proportion of homogenized medium was larger compared to the structurally resolved electrode. For $S_1<15\,\%$, it behaves similar to the pure electrode, but is shifted by $S_1>0$. This similarity is because the electrolyte had not yet reached the separator, but was influenced by the pure electrode only. At larger saturations, a pressure plateau is reached which was also observed in the literature \cite{Tavangarrad2019,Sauter2020}. Here, it corresponds to the smooth filling of the fully homogenized separator. Along this plateau, the pressure increased by 0.4\,kPa only, while at the end of the filling, a large pressure increase was observed due to the electrolyte imbibition into the small pores of the electrode. However, a notable gas entrapment remains. Note that in general, the shift along $S_1>0$ for the homogenized microstructures was also influenced by the definition of the saturation (cf.\ \autoref{eq:Saturation}), where adding lattice cells with a bounce-back fraction $n_\mathrm{s} \neq 0$ led to a reduction of the total pore space.

\autoref{fig:pS}b)$-$d) show snapshots of simulation cross sections ($xy$-plane) at $t=400{,}000$\,ts for the three different microstructures. The corresponding state points on the pressure-saturation curves are indicated by the hashes in \autoref{fig:pS}a). Compared to the pure electrode (cf.\ \autoref{fig:pS}b) ), in \autoref{fig:pS}c) slightly more electrolyte is imbibed into pores with binder. \autoref{fig:pS}d) shows the electrolyte distribution in the resolved electrode and the homogenized separator. While the electrode part is not yet completely filled, a breakthrough of the electrolyte into the separator is observed. This state corresponds to the pressure plateau shown in \autoref{fig:pS}a). Moreover, \autoref{fig:pS}d) indicates some interesting model-related phenomena. On the one hand, the heterogeneity of $n_\mathrm{s}$ values in the separator can be used to mimic structural effects that influence gas entrapment, the flow orientation and deformation of the electrolyte-gas interface. On the other hand, and even more importantly, the interface thickness in the homogenized separator is constant and not affected by the distribution of $n_\mathrm{s}$. It is also identical to the interface thickness in the structurally resolved pores for which $n_\mathrm{s}=0$.

These simulations give evidence for the potential and strength of the HMCSC. Moreover, they demonstrate possible applications of multi-phase flows in heterogeneous porous media going beyond electrochemical energy storage showcased in this work. 

\section{Conclusion}  \label{sec:Conclusion}
A new lattice Boltzmann model was presented which is especially useful for the simulation of multi-phase flows in heterogeneous porous media, but can also be used to study single-phase flows. It follows the approach of Pereira \cite{Pereira2016} and combines the grayscale method of Walsh~\textit{et al.} (GS-WBS) \cite{Walsh2009}, i.e.\ a homogenization approach, with the multi-component Shan-Chen model (MCSC) \cite{Shan1993}. Therefore, it was called the homogenized multi-component Shan-Chen method (HMCSC).

The HMCSC was tested using a broad variety of benchmark scenarios that are typically used to validate single- and two-phase flow phenomena in porous media. The results were compared to analytical and semi-analytical solutions where available and shown to agree well. In addition, the HMCSC was applied to study the electrolyte filling of different variants of a realistic 3D reconstruction of a lithium-ion battery microstructure. Hereby, the new method was shown to reproduce the relevant physical phenomena also within the homogenized nanoporous binder and separator. For example, including the wetting binder led to a reduction of the capillary pressure and an improvement of the final degree of saturation which is beneficial for the battery performance. Moreover, when adding the fully homogenized separator with space-dependent $n_\mathrm{s}$ values, structural effects which influence gas entrapment, characteristic flow paths, and the deformation of the electrolyte-gas interface were mimicked appropriately. 

The HMCSC brings together the physical characteristics of GS-WBS and MCSC, and additionally overcomes some deficiencies of GS-WBS as well as the model reported by Pereira \cite{Pereira2016}. In contrast to the latter, the HMCSC is consistent with the MCSC. The MCSC-related model parameters $G_{\mathrm{ads}}$ and $G_{\mathrm{inter}}$ can be chosen identically to the parameters in the original MCSC (cf.\ \cite{Huang2007}). The corresponding physical properties, i.e. contact angle and interfacial tension, are not affected by the homogenization. They are constant and properties of the fluid only, even in heterogeneous porous media where $n_\mathrm{s}$ is space-dependent. Thus, no further parametrization is required. Moreover, in contrast to the GS-WBS, no artificial force and velocity discontinuity was observed at interfaces between different porous media. We assume that this is due to the Shan-Chen forcing scheme that was used for the HMCSC.

Altogether, the HMCSC is physically motivated, accurate, and efficient. Nonetheless, it is simple, intuitive, easy to implement and allows a straightforward parametrization that is consistent with the original MCSC. The results are promising. By the right choice of model parameters, it can be applied to all situations for which either the original MCSC or the GS-WBS can be used. However, the HMCSC is not restricted to these particular applications. All other research fields where multi-phase fluid flow occurs in heterogeneous and multi-scale porous media can benefit from this method, too. Potential applications are subsoil or groundwater flow \cite{Pereira2016,Ning2019}, oil recovery via water injection \cite{Spaid1998,Welge1952}, biofilm growth in porous structures \cite{Jung2021}, the characterization of gas diffusion electrodes \cite{Danner2016}, water transport in gas diffusion layers \cite{Zhu2021}, or electrolyte filling of batteries \cite{Shodiev2021,Lautenschlaeger2022}.

\section*{Acknowledgement}  \label{sec:Acknowledgement}
The authors gratefully acknowledge financial support from the European Union's Horizon 2020 Research and Innovation Programme within the project ``DEFACTO'' under the grant number 875247. The simulations were carried out on the Hawk at the High Performance Computing Center Stuttgart (HLRS) under the grant LaBoRESys, and on JUSTUS~2 at the University Ulm under the grant INST 40/467-1 FUGG.

\clearpage
\begin{appendices}
\renewcommand{\appendixautorefname}{Section}
\renewcommand{\theequation}{(\thesection.\arabic{equation})}
\setcounter{equation}{0}
\renewcommand{\thetable}{\thesection.\arabic{table}}
\setcounter{table}{0}
\renewcommand{\thefigure}{\thesection.\arabic{figure}}
\setcounter{figure}{0}
\section{Lattice Boltzmann method} \label{sec:Appendix_LBM}
\noindent LBM solves the discretized Boltzmann equation 
\begin{align}
    \label{eq:LBGK}
    \begin{split}
    f_{i}\left(\mathbf{x}+\mathbf{c}_i \Delta t, t + \Delta t \right) =  f_{i}\left(\mathbf{x}, t\right) - \frac{\Delta t}{\tau} \left(f_{i}\left(\mathbf{x}, t\right) - f^{\mathrm{eq}}_{i}\left(\mathbf{x}, t\right)\right)
    \end{split},
\end{align}
using discrete distribution functions $f_i$, also referred to as populations. This approach uses the BGK collision \cite{Bhatnagar1954}, which describes the relaxation of $\boldsymbol{f}$ towards the Maxwell-Boltzmann equilibrium distribution function $\boldsymbol{f}^\mathrm{eq}$ (cf. \autoref{eq:Maxwell}) with a characteristic relaxation time $\tau$. The relaxation time determines the kinematic viscosity via $\nu=c_\mathrm{s}^{2}(\tau-1/2\Delta t)$, where $c_\mathrm{s}^{2}$ is the velocity set dependent lattice speed of sound.

The discretization of \autoref{eq:LBGK} is done in velocity space on a regular square or cubic lattice, for 2D and 3D, respectively. Each lattice cell is linked to its adjacent neighbors, denoted with $i$. The resulting velocity sets, $\mathbf{c}$, are called D2Q9 and D3Q27, which is often reduced to D3Q19 for efficiency, and are given in \eqsref{eq:D2Q9}{eq:D3Q19}. For completeness, $t$ and $\mathbf{x}$ denote time and lattice location, with $\Delta t$ and $\Delta \mathrm{x}$ being the temporal and spatial step. For the non-dimensional computationally efficient form, $\Delta \mathrm{x}$ and $\Delta t$ are unity. As such they are usually omitted, but are included in this work for clarity and dimensional consistency.

The homogenization approach by Walsh~\textit{et al.} \cite{Walsh2009} is
\begin{align}
    \label{eq:LBGKWBS}
    \begin{split}
    f_{i}\left(\mathbf{x}+\mathbf{c}_i \Delta t, t + \Delta t \right) =&\quad\  (1-n_{\mathrm{s}}(\mathbf{x}))f_{i}\left(\mathbf{x}, t\right) \\
    &- (1-n_{\mathrm{s}}(\mathbf{x}))\frac{\Delta t}{\tau} \left(f_{i}\left(\mathbf{x}, t\right) - f^{\mathrm{eq}}_{i}\left(\mathbf{x}, t\right)\right) \\
    &+ n_{\mathrm{s}}(\mathbf{x}) f_{\bar{i}}\left(\mathbf{x}, t^{\ast}\right).
    \end{split}
\end{align}
Here, $n_\mathrm{s}\in[0,1]$ acts as an interpolation factor that scales the fluid and solid behavior of $\boldsymbol{f}$. From all populations $\boldsymbol{f}$ in a lattice cell, the fraction $(1-n_\mathrm{s})$ is allowed to flow freely and therefore behaves fluid-like (cf. first and the second line of \autoref{eq:LBGKWBS}), while the fraction $n_\mathrm{s}$ is bounced back and therefore behaves solid-like (cf. third line of \autoref{eq:LBGKWBS}). The latter term uses the pre-collision populations which is highlighted by $t^{\ast}$. The symbol $\bar{i}$ denotes the direction opposite to $i$ with the exception $i=0=\bar{i}$. Note that for $n_\mathrm{s}=0$, \autoref{eq:LBGKWBS} is equivalent to the default LB BGK equation that describes pure fluid flow, \autoref{eq:LBGK}, while for $n_\mathrm{s}=1$ it corresponds to pure bounce-back that describes a no-slip wall. 

The MCSC approach has been fully covered in \autoref{sec:Model}. It is iterated that for each component $\sigma$ used, a unique discrete population function $\boldsymbol{f}_\sigma$ is needed. Therefore, in the following $\sigma$ is explicitly mentioned in all equations.

\subsection{Maxwell-Boltzmann distribution}
The Maxwell-Boltzmann equilibrium distribution function is given by
\begin{align}
    \label{eq:Maxwell}
    f^{\mathrm{eq}}_{i,\sigma} = w_i \rho_{\sigma} \left[1 + \frac{\mathbf{c}_i \mathbf{u}^{\mathrm{eq}}_{\sigma}}{c_\mathrm{s}^2} + \frac{(\mathbf{c}_i \mathbf{u}^{\mathrm{eq}}_{\sigma})^2}{2 c_\mathrm{s}^4} - \frac{\mathbf{u}^{\mathrm{eq}}_{\sigma} \mathbf{u}^{\mathrm{eq}}_{\sigma}}{2 c_\mathrm{s}^2}\right].
\end{align}
Here, $w_i$ are the lattice specific weights, $\rho_{\sigma}$ is the fluid density of component $\sigma$ (cf. \autoref{eq:rho_BGK}) and $\mathbf{u}^{\mathrm{eq}}$ is the equilibrium velocity (cf. \autoref{eq:SC_u_eq}).

\subsection{Multi-phase physical quantities}
The macroscopic physical quantities can be recovered from the discrete distribution function. In the following they are expressed in terms of multiple phases, but apply for single-phase fluids as well.

The fluid density of component $\sigma$ is the zeroth moment of $\boldsymbol{f}_{\sigma}$ and determined as
\begin{align}
    \label{eq:rho_BGK}
    \rho_{\sigma} = \sum_{i}^{}f_{i,\sigma}.
\end{align}

The velocity of component $\sigma$ is the first moment of $\boldsymbol{f}_{\sigma}$ and determined as
\begin{align}
    \label{eq:mom_BGK}
    \mathbf{u}_{\sigma} = \frac{1}{\rho_{\sigma}}\sum_{i}^{}f_{i,\sigma}\textbf{c}_i.
\end{align}

The total pressure $p$ of the mixture follows the ideal gas law. It is extended by a contribution from the fluid-fluid interaction which is especially relevant at the interface. $p$ is determined as
\begin{align}    
    \label{eq:SC_pressure}
    p(\mathbf{x}) = c_s^2 \left[\rho(\mathbf{x}) + G_{\mathrm{inter},\sigma\bar{\sigma}}\rho_{\sigma}(\mathbf{x})\rho_{\bar{\sigma}}(\mathbf{x}) \Delta t^2\right].
\end{align}

\subsection{Velocity Sets}
The HMCSC has been implemented for both 2D and 3D simulations. The corresponding velocity sets used for the simulations of the present work, i.e.\ D2Q9 and D3Q19, are given in the following:

\noindent\textbf{D2Q9}
\begin{align}
    \label{eq:D2Q9}
    &{\left[\mathbf{c}_{0}, \mathbf{c}_{1}, \mathbf{c}_{2}, \mathbf{c}_{3}, \mathbf{c}_{4}, \mathbf{c}_{5}, \mathbf{c}_{6}, \mathbf{c}_{7}, \mathbf{c}_{8}\right]} \\
    &\quad=\frac{\Delta x}{\Delta t}\left[\begin{array}{rrrrrrrrr}
        0 & -1 & -1 & -1 &  0 &  1 & 1 & 1 & 0 \\
        0 &  1 &  0 & -1 & -1 & -1 & 0 & 1 & 1
    \end{array}\right]
   \nonumber
\end{align}
The D2Q9 weights are: $w_i=4/9$ for $|\mathbf{c}_i|=0$, $w_i=1/9$ for $|\mathbf{c}_i|=1$, and $w_i=1/36$ for $|\mathbf{c}_i|=\sqrt{2}$ with the speed of sound $c_\mathrm{s}=1/\sqrt{3} \Delta \mathbf{x}/\Delta t$.

\vspace{2mm}\noindent\textbf{D3Q19}
\begin{align}
    \label{eq:D3Q19}
    &{\left[\mathbf{c}_{0}, \mathbf{c}_{1}, \mathbf{c}_{2}, \mathbf{c}_{3}, \mathbf{c}_{4}, \mathbf{c}_{5}, \mathbf{c}_{6}, \mathbf{c}_{7}, \mathbf{c}_{8}, \mathbf{c}_{9}, \mathbf{c}_{10}, \mathbf{c}_{11}, \mathbf{c}_{12}, \mathbf{c}_{13}, \mathbf{c}_{14}, \mathbf{c}_{15}, \mathbf{c}_{16}, \mathbf{c}_{17}, \mathbf{c}_{18}\right]} \\
    &\quad=\frac{\Delta x}{\Delta t}\left[\begin{array}{rrrrrrrrrrrrrrrrrrr}
    0 & -1 &  0 &  0 & -1 & -1 & -1 & -1 &  0 &  0 &  1 &  0 &  0 &  1 &  1 &  1 &  1 & 0 &  0 \\
    0 &  0 & -1 &  0 & -1 &  1 &  0 &  0 & -1 & -1 &  0 &  1 &  0 &  1 & -1 &  0 &  0 & 1 &  1 \\
    0 &  0 &  0 & -1 &  0 &  0 & -1 &  1 & -1 &  1 &  0 &  0 &  1 &  0 &  0 &  1 & -1 & 1 & -1
    \end{array}\right]
   \nonumber
\end{align}
The D3Q19 weights are: $w_i=1/3$ for $|\mathbf{c}_i|=0$, $w_i=1/18$ for $|\mathbf{c}_i|=1$, and $w_i=1/36$ for $|\mathbf{c}_i|=\sqrt{2}$ with the speed of sound $c_\mathrm{s}=1/\sqrt{3} \Delta \mathbf{x}/\Delta t$.

\section{Guo vs. Shan-Chen Forcing} \label{sec:Appendix_GuoShan}
It has been reported in literature \cite{Zhu2013,Yehya2015,Ginzburg2016} and also mentioned in the main text of this paper, that using the GS-WBS can result in velocity discontinuities at interfaces between different porous media. This was not observed when using the HMCSC (cf.\ \autoref{fig:DarcyBrinkmanGinzburg}). The reason might be due to the fact that the HMCSC uses the Shan-Chen forcing scheme \cite{Shan1993}, while the Guo forcing scheme \cite{Guo2002a} was used for the simulations reported in \cite{Zhu2013,Ginzburg2016}. 

The differences between both forcing schemes are briefly outlined in the following. For brevity regarding the notation, the position $\mathbf{x}$ is omitted and the time step is indicated in the exponent. The meaning of which is: $t$ is the current time step, and $\tilde{t}$ denotes the state just before streaming. To emphasize and avoid confusion, $t^{\ast}$ is used in addition to denote the pre-collision state.
\begin{align}
    \label{eq:GS-WBS}
    \textbf{GS-WBS:~~}& 
    f_{i}^{\tilde{t}} =& (1-n_{\mathrm{s}})f_{i}^{t} - (1-n_{\mathrm{s}})\frac{\Delta t}{\tau} \left(f_{i}^{t} - f^{\mathrm{eq},t}_{i}(\textbf{u}_\mathrm{macro})\right) + n_{\mathrm{s}} f_{\bar{i}}^{t^{\ast}}\\ &&+
    (1-n_{\mathrm{s}})F_i(\boldsymbol{F}_\mathrm{tot},\textbf{u}_\mathrm{macro})\nonumber\\
    \label{eq:HMCSC}
    \textbf{HMCSC:~~}&
    f_{i}^{\tilde{t}} =& (1-n_{\mathrm{s}})f_{i}^{t} - (1-n_{\mathrm{s}})\frac{\Delta t}{\tau} \left(f_{i}^{t} - f^{\mathrm{eq},t}_{i}(\textbf{u}^{\mathrm{eq}})\right) + n_{\mathrm{s}} f_{\bar{i}}^{t^{\ast}}
\end{align}
\eqsref{eq:GS-WBS}{eq:HMCSC} show the GS-type LB BGK equation in the Guo and Shan-Chen forcing scheme, respectively. The most obvious difference between both is the force term $F_i$ in \autoref{eq:GS-WBS}. It is directly dependent on the total force $\boldsymbol{F}_\mathrm{tot}$. The whole term is scaled by the factor $(1-n_\mathrm{s})$. This means that also the fluid-fluid and solid-fluid force contributions are scaled by $(1-n_\mathrm{s})$ leading to an $n_\mathrm{s}$-dependent interfacial tension and wetting behavior which was shown to be circumvented when using the HMCSC. Moreover, at the interface between two porous media from which the $n_\mathrm{s}$ values differ, the force term is also scaled differently, which might be one reason for the force and velocity discontinuity observed in the literature \cite{Zhu2013,Yehya2015,Ginzburg2016}.

Another might be that $\boldsymbol{F}_\mathrm{tot}$ enters also indirectly into \autoref{eq:GS-WBS} via $\textbf{u}_\mathrm{macro}$ which in the Guo forcing scheme is needed for the calculation of $\boldsymbol{f}^{\mathrm{eq}}$. However, as $\textbf{u}_\mathrm{macro}$ itself is already scaled by $(1-n_\mathrm{s})$ (cf. \autoref{eq:HMCSC_u_macro}), the factor $(1-n_\mathrm{s})$ might be unintentionally considered several times for the force calculation.

Now, considering the Shan-Chen forcing scheme, the situation is much clearer. $\boldsymbol{F}_\mathrm{tot}$ is only indirectly incorporated into $\textbf{u}^{\mathrm{eq}}$ (cf. \autoref{eq:SC_u_eq}) which in the Shan-Chen forcing scheme is needed for the calculation of $\boldsymbol{f}^{\mathrm{eq}}$. Thus, a repeated scaling by $(1-n_\mathrm{s})$ is impossible.

Nevertheless, the actual reason why GS-WBS in Guo forcing leads to velocity discontinuities, while it does not when using the HMCSC in the Shan-Chen forcing scheme remains still unclear. A detailed analysis would be necessary which, however, goes far beyond the scope of this study.

\section{Data} \label{sec:Data}
\subsection{Maximum velocity in Open Boundary Flow}
The maximum velocities in lattice units for the open boundary flow validation test are presented in \autoref{tab:DarcyBrinkmanGinzburg}.
\begin{table}[!ht]
	\centering
	\caption{Numerical values of the maximum velocities $u_{\mathrm{max}}$ that were used to normalize the velocity profiles in \autoref{fig:DarcyBrinkmanGinzburg}. }
	\begin{tabular}{l||l}
		\toprule \toprule
        $n_{\mathrm{s},2}$ & $u_{\mathrm{max}}$ [lu/ts]\\
        \midrule
		0.001 & 4.732 $\cdot 10^{-4}$\\
		0.01  & 4.948 $\cdot 10^{-5}$\\
		0.1   & 4.500 $\cdot 10^{-6}$\\
		0.5   & 5.000 $\cdot 10^{-7}$\\
		0.8   & 1.250 $\cdot 10^{-7}$\\
		\bottomrule \bottomrule 
	\end{tabular}%
	\label{tab:DarcyBrinkmanGinzburg}%
\end{table}%

\subsection{Structure Generation of Homogenized Separator} 
The homogenized separator structure has been created by downsampling the original binarized image by a factor of 20 in each direction. The original structure consisted of $600\times820\times560$ voxels with a voxel length of $2.19\cdot 10^{-8}$\,m. This structure has been subdivided into $20\times20\times20$ cubes, with the volume fraction of each cube being the mean volume fraction of its containing voxels. The resulting structure is of shape $30\times41\times28$. The voxel length, i.e.\ $4.38\cdot 10^{-7}$\,m, is identical to that of voxels of the electrode structure. But since the electrode structure is much larger, the separator has been stacked multiple times in both directions perpendicular to the filling direction to fit the dimensions of the electrode.

This procedure seems trivial and one could get the impression that it would be possible to run simulations using the gray values from CT images directly. However, as discussed in Ref.\,\cite{Baveye2017}, this is not the case. The model parameter called the bounce-back fraction $n_\mathrm{s}$ does not correspond to the physical bounce-back fraction that is determined through images. With no information about the actual penetrability of the material, the relation between the physical parameter and the model parameter can only be guessed. To overcome this limitation, investigating a method for converting both will be one topic of our future research.

\end{appendices}
\unappendix
\clearpage

\bibliographystyle{elsarticle-num} 

\bibliography{main}

\begin{thebibliography}{10}
\expandafter\ifx\csname url\endcsname\relax
  \def\url#1{\texttt{#1}}\fi
\expandafter\ifx\csname urlprefix\endcsname\relax\def\urlprefix{URL }\fi
\expandafter\ifx\csname href\endcsname\relax
  \def\href#1#2{#2} \def\path#1{#1}\fi

\bibitem{Kang2007}
Q.~Kang, P.~C. Lichtner, D.~Zhang, {An improved lattice Boltzmann model for multicomponent reactive transport in porous media at the pore scale}, Water Resources Research 43~(12) (2007) 1--12.
\newblock \href {https://doi.org/10.1029/2006WR005551} {\path{doi:10.1029/2006WR005551}}.

\bibitem{Dentz2011}
M.~Dentz, T.~{Le Borgne}, A.~Englert, B.~Bijeljic, {Mixing, spreading and reaction in heterogeneous media: A brief review}, Journal of Contaminant Hydrology 120-121~(C) (2011) 1--17.
\newblock \href {https://doi.org/10.1016/j.jconhyd.2010.05.002} {\path{doi:10.1016/j.jconhyd.2010.05.002}}.

\bibitem{Steefel2005}
C.~Steefel, D.~DePaolo, P.~Lichtner, {Reactive transport modeling: An essential tool and a new research approach for the Earth sciences}, Earth and Planetary Science Letters 240~(3-4) (2005) 539--558.
\newblock \href {https://doi.org/10.1016/j.epsl.2005.09.017} {\path{doi:10.1016/j.epsl.2005.09.017}}.

\bibitem{Baveye2017}
P.~C. Baveye, V.~Pot, P.~Garnier, {Accounting for sub-resolution pores in models of water and solute transport in soils based on computed tomography images: Are we there yet?}, Journal of Hydrology 555 (2017) 253--256.
\newblock \href {https://doi.org/10.1016/j.jhydrol.2017.10.021} {\path{doi:10.1016/j.jhydrol.2017.10.021}}.

\bibitem{Laubach2019}
S.~E. Laubach, R.~H. Lander, L.~J. Criscenti, L.~M. Anovitz, J.~L. Urai, R.~M. Pollyea, J.~N. Hooker, W.~Narr, M.~A. Evans, S.~N. Kerisit, J.~E. Olson, T.~Dewers, D.~Fisher, R.~Bodnar, B.~Evans, P.~Dove, L.~M. Bonnell, M.~P. Marder, L.~Pyrak‐Nolte, {The Role of Chemistry in Fracture Pattern Development and Opportunities to Advance Interpretations of Geological Materials}, Reviews of Geophysics 57~(3) (2019) 1065--1111.
\newblock \href {https://doi.org/10.1029/2019RG000671} {\path{doi:10.1029/2019RG000671}}.

\bibitem{Yuan2019}
G.~Yuan, Y.~Cao, H.-M. Schulz, F.~Hao, J.~Gluyas, K.~Liu, T.~Yang, Y.~Wang, K.~Xi, F.~Li, {A review of feldspar alteration and its geological significance in sedimentary basins: From shallow aquifers to deep hydrocarbon reservoirs}, Earth-Science Reviews 191~(October 2017) (2019) 114--140.
\newblock \href {https://doi.org/10.1016/j.earscirev.2019.02.004} {\path{doi:10.1016/j.earscirev.2019.02.004}}.

\bibitem{Blunt2013}
M.~J. Blunt, B.~Bijeljic, H.~Dong, O.~Gharbi, S.~Iglauer, P.~Mostaghimi, A.~Paluszny, C.~Pentland, {Pore-scale imaging and modelling}, Advances in Water Resources 51 (2013) 197--216.
\newblock \href {https://doi.org/10.1016/j.advwatres.2012.03.003} {\path{doi:10.1016/j.advwatres.2012.03.003}}.

\bibitem{Zhang2020}
Q.~Zhang, H.~Yu, X.~Li, T.~Liu, J.~Hu, {A New Upscaling Method for Fluid Flow Simulation in Highly Heterogeneous Unconventional Reservoirs}, Geofluids 2020 (2020) 1--11.
\newblock \href {https://doi.org/10.1155/2020/6213183} {\path{doi:10.1155/2020/6213183}}.

\bibitem{Sok2010}
R.~M. Sok, T.~Varslot, A.~Ghous, S.~Latham, A.~P. Sheppard, M.~A. Knackstedt, {Pore scale characterization of carbonates at multiple scales: Integration of micro-CT, BSEM, and FIBSEM}, Petrophysics 51~(6) (2010) 379--387.

\bibitem{Bai2013}
B.~Bai, M.~Elgmati, H.~Zhang, M.~Wei, {Rock characterization of Fayetteville shale gas plays}, Fuel 105 (2013) 645--652.
\newblock \href {https://doi.org/10.1016/j.fuel.2012.09.043} {\path{doi:10.1016/j.fuel.2012.09.043}}.

\bibitem{Zhang2016}
X.~Zhang, J.~W. Crawford, R.~J. Flavel, I.~M. Young, {A multi-scale Lattice Boltzmann model for simulating solute transport in 3D X-ray micro-tomography images of aggregated porous materials}, Journal of Hydrology 541 (2016) 1020--1029.
\newblock \href {https://doi.org/10.1016/j.jhydrol.2016.08.013} {\path{doi:10.1016/j.jhydrol.2016.08.013}}.

\bibitem{Kang2019}
D.~H. Kang, E.~Yang, T.~S. Yun, {Stokes‐Brinkman Flow Simulation Based on 3‐D $\mu$‐CT Images of Porous Rock Using Grayscale Pore Voxel Permeability}, Water Resources Research 55~(5) (2019) 4448--4464.
\newblock \href {https://doi.org/10.1029/2018WR024179} {\path{doi:10.1029/2018WR024179}}.

\bibitem{Soulaine2019}
C.~Soulaine, P.~Creux, H.~A. Tchelepi, {Micro-continuum Framework for Pore-Scale Multiphase Fluid Transport in Shale Formations}, Transport in Porous Media 127~(1) (2019) 85--112.
\newblock \href {https://doi.org/10.1007/s11242-018-1181-4} {\path{doi:10.1007/s11242-018-1181-4}}.

\bibitem{Mehmani2020}
A.~Mehmani, R.~Verma, M.~Prodanovi{\'{c}}, {Pore-scale modeling of carbonates}, Marine and Petroleum Geology 114 (2020) 104141.
\newblock \href {https://doi.org/10.1016/j.marpetgeo.2019.104141} {\path{doi:10.1016/j.marpetgeo.2019.104141}}.

\bibitem{Kang2002}
Q.~Kang, D.~Zhang, S.~Chen, {Unified lattice Boltzmann method for flow in multiscale porous media}, Physical Review E 66~(5) (2002) 056307.
\newblock \href {https://doi.org/10.1103/PhysRevE.66.056307} {\path{doi:10.1103/PhysRevE.66.056307}}.

\bibitem{Krueger2016}
T.~Krueger, H.~Kusumaatmaja, A.~Kuzmin, O.~Shardt, G.~Silva, E.~Viggen, The Lattice {B}oltzmann Method: Principles and Practice, Springer, 2016.

\bibitem{Liu2016}
H.~Liu, Q.~Kang, C.~R. Leonardi, S.~Schmieschek, A.~Narv{\'{a}}ez, B.~D. Jones, J.~R. Williams, A.~J. Valocchi, J.~Harting, Multiphase lattice {B}oltzmann simulations for porous media applications: A review, Computational Geosciences 20~(4) (2016) 777--805.
\newblock \href {https://doi.org/10.1007/s10596-015-9542-3} {\path{doi:10.1007/s10596-015-9542-3}}.

\bibitem{Chen2014}
L.~Chen, Q.~Kang, Y.~Mu, Y.~L. He, W.~Q. Tao, A critical review of the pseudopotential multiphase lattice {B}oltzmann model: Methods and applications, International Journal of Heat and Mass Transfer 76 (2014) 210--236.
\newblock \href {https://doi.org/10.1016/j.ijheatmasstransfer.2014.04.032} {\path{doi:10.1016/j.ijheatmasstransfer.2014.04.032}}.

\bibitem{Spaid1997}
M.~A.~A. Spaid, F.~R. Phelan, {Lattice Boltzmann methods for modeling microscale flow in fibrous porous media}, Physics of Fluids 9~(9) (1997) 2468--2474.
\newblock \href {https://doi.org/10.1063/1.869392} {\path{doi:10.1063/1.869392}}.

\bibitem{Freed1998}
D.~M. Freed, {Lattice-Boltzmann Method for Macroscopic Porous Media Modeling}, International Journal of Modern Physics C 09~(08) (1998) 1491--1503.
\newblock \href {https://doi.org/10.1142/S0129183198001357} {\path{doi:10.1142/S0129183198001357}}.

\bibitem{Guo2002}
Z.~Guo, T.~S. Zhao, {Lattice Boltzmann model for incompressible flows through porous media}, Physical Review E 66~(3) (2002) 036304.
\newblock \href {https://doi.org/10.1103/PhysRevE.66.036304} {\path{doi:10.1103/PhysRevE.66.036304}}.

\bibitem{Ginzburg2008}
I.~Ginzburg, {Consistent lattice Boltzmann schemes for the Brinkman model of porous flow and infinite Chapman-Enskog expansion}, Physical Review E 77~(6) (2008) 066704.
\newblock \href {https://doi.org/10.1103/PhysRevE.77.066704} {\path{doi:10.1103/PhysRevE.77.066704}}.

\bibitem{Gao2014}
J.~Gao, H.~Xing, Z.~Tian, H.~Muhlhaus, {Lattice Boltzmann modeling and evaluation of fluid flow in heterogeneous porous media involving multiple matrix constituents}, Computers {\&} Geosciences 62 (2014) 198--207.
\newblock \href {https://doi.org/10.1016/j.cageo.2013.07.019} {\path{doi:10.1016/j.cageo.2013.07.019}}.

\bibitem{Ginzburg2015}
I.~Ginzburg, G.~Silva, L.~Talon, {Analysis and improvement of Brinkman lattice Boltzmann schemes: Bulk, boundary, interface. similarity and distinctness with finite elements in heterogeneous porous media}, Physical Review E 91~(2) (2015) 1--32.
\newblock \href {https://doi.org/10.1103/PhysRevE.91.023307} {\path{doi:10.1103/PhysRevE.91.023307}}.

\bibitem{Dardis1998}
O.~Dardis, J.~McCloskey, {Lattice Boltzmann scheme with real numbered solid density for the simulation of flow in porous media}, Physical Review E 57~(4) (1998) 4834--4837.
\newblock \href {https://doi.org/10.1103/PhysRevE.57.4834} {\path{doi:10.1103/PhysRevE.57.4834}}.

\bibitem{Thorne2004}
D.~Thorne, M.~Sukop, Lattice boltzmann model for the elder problem, in: C.~T. Miller, G.~F. Pinder (Eds.), Computational Methods in Water Resources: Volume 2, Vol.~55 of Developments in Water Science, Elsevier, 2004, pp. 1549--1557.
\newblock \href {https://doi.org/https://doi.org/10.1016/S0167-5648(04)80165-5} {\path{doi:https://doi.org/10.1016/S0167-5648(04)80165-5}}.

\bibitem{Chen2008}
Y.~Chen, K.~Zhu, {A study of the upper limit of solid scatters density for gray Lattice Boltzmann Method}, Acta Mechanica Sinica 24~(5) (2008) 515--522.
\newblock \href {https://doi.org/10.1007/s10409-008-0167-9} {\path{doi:10.1007/s10409-008-0167-9}}.

\bibitem{Walsh2009}
S.~D. Walsh, H.~Burwinkle, M.~O. Saar, A new partial-bounceback lattice {B}oltzmann method for fluid flow through heterogeneous media, Computers {\&} Geosciences 35~(6) (2009) 1186--1193.
\newblock \href {https://doi.org/10.1016/j.cageo.2008.05.004} {\path{doi:10.1016/j.cageo.2008.05.004}}.

\bibitem{Zhu2013}
J.~Zhu, J.~Ma, {An improved gray lattice Boltzmann model for simulating fluid flow in multi-scale porous media}, Advances in Water Resources 56 (2013) 61--76.
\newblock \href {https://doi.org/10.1016/j.advwatres.2013.03.001} {\path{doi:10.1016/j.advwatres.2013.03.001}}.

\bibitem{Yoshida2014}
H.~Yoshida, H.~Hayashi, {Transmission-Reflection Coefficient in the Lattice Boltzmann Method}, Journal of Statistical Physics 155~(2) (2014) 277--299.
\newblock \href {https://doi.org/10.1007/s10955-014-0953-7} {\path{doi:10.1007/s10955-014-0953-7}}.

\bibitem{Yehya2015}
A.~Yehya, H.~Naji, M.~C. Sukop, {Simulating flows in multi-layered and spatially-variable permeability media via a new Gray Lattice Boltzmann model}, Computers and Geotechnics 70 (2015) 150--158.
\newblock \href {https://doi.org/10.1016/j.compgeo.2015.07.017} {\path{doi:10.1016/j.compgeo.2015.07.017}}.

\bibitem{Ginzburg2016}
I.~Ginzburg, {Comment on “An improved gray Lattice Boltzmann model for simulating fluid flow in multi-scale porous media”: Intrinsic links between LBE Brinkman schemes}, Advances in Water Resources 88 (2016) 241--249.
\newblock \href {https://doi.org/10.1016/j.advwatres.2014.05.007} {\path{doi:10.1016/j.advwatres.2014.05.007}}.

\bibitem{Zhu2018}
J.~Zhu, J.~Ma, Extending a gray lattice {B}oltzmann model for simulating fluid flow in multi-scale porous media, Scientific Reports 8~(1) (2018) 1--19.
\newblock \href {https://doi.org/10.1038/s41598-018-24151-2} {\path{doi:10.1038/s41598-018-24151-2}}.

\bibitem{McDonald2016}
P.~J. McDonald, M.~N. Turner, Combining effective media and multi-phase methods of lattice {B}oltzmann modelling for the characterisation of liquid-vapour dynamics in multi-length scale heterogeneous structural materials, Modelling and Simulation in Materials Science and Engineering 24~(1) (2016) 015010.
\newblock \href {https://doi.org/10.1088/0965-0393/24/1/015010} {\path{doi:10.1088/0965-0393/24/1/015010}}.

\bibitem{Zalzale2016}
M.~Zalzale, M.~Ramaioli, K.~L. Scrivener, P.~J. McDonald, {Gray free-energy multiphase lattice Boltzmann model with effective transport and wetting properties}, Physical Review E 94~(5) (2016) 1--13.
\newblock \href {https://doi.org/10.1103/PhysRevE.94.053301} {\path{doi:10.1103/PhysRevE.94.053301}}.

\bibitem{Lei2019}
S.~Lei, Y.~Shi, {Separate-phase model and its lattice Boltzmann algorithm for liquid-vapor two-phase flows in porous media}, Physical Review E 99~(5) (2019) 053302.
\newblock \href {https://doi.org/10.1103/PhysRevE.99.053302} {\path{doi:10.1103/PhysRevE.99.053302}}.

\bibitem{Ning2019}
Y.~Ning, C.~Wei, G.~Qin, {A unified grayscale lattice Boltzmann model for multiphase fluid flow in vuggy carbonates}, Advances in Water Resources 124 (2019) 68--83.
\newblock \href {https://doi.org/10.1016/j.advwatres.2018.12.007} {\path{doi:10.1016/j.advwatres.2018.12.007}}.

\bibitem{An2020}
S.~An, Y.~Zhan, J.~Yao, H.~W. Yu, V.~Niasar, {A greyscale volumetric lattice Boltzmann method for upscaling pore-scale two-phase flow}, Advances in Water Resources 144 (2020) 103711.
\newblock \href {https://doi.org/10.1016/j.advwatres.2020.103711} {\path{doi:10.1016/j.advwatres.2020.103711}}.

\bibitem{Pereira2016}
G.~G. Pereira, Grayscale lattice {B}oltzmann model for multiphase heterogeneous flow through porous media, Physical Review E 93~(6) (2016) 1--14.
\newblock \href {https://doi.org/10.1103/PhysRevE.93.063301} {\path{doi:10.1103/PhysRevE.93.063301}}.

\bibitem{Shan1993}
X.~Shan, H.~Chen, {L}attice {B}oltzmann model for simulating flows with multiple phases and components, Physical Review E 47~(3) (1993) 1815--1819.
\newblock \href {https://doi.org/10.1103/PhysRevE.47.1815} {\path{doi:10.1103/PhysRevE.47.1815}}.

\bibitem{Ghezzehei2012}
T.~A. Ghezzehei, {Linking sub-pore scale heterogeneity of biological and geochemical deposits with changes in permeability}, Advances in Water Resources 39 (2012) 1--6.
\newblock \href {https://doi.org/10.1016/j.advwatres.2011.12.015} {\path{doi:10.1016/j.advwatres.2011.12.015}}.

\bibitem{Hassannayebi2021}
N.~Hassannayebi, B.~Jammernegg, J.~Schritter, P.~Arnold, F.~Enzmann, M.~Kersten, A.~P. Loibner, M.~Fern{\o}, H.~Ott, {Relationship Between Microbial Growth and Hydraulic Properties at the Sub-Pore Scale}, Transport in Porous Media 139~(3) (2021) 579--593.
\newblock \href {https://doi.org/10.1007/s11242-021-01680-5} {\path{doi:10.1007/s11242-021-01680-5}}.

\bibitem{Krevor2012}
S.~C.~M. Krevor, R.~Pini, L.~Zuo, S.~M. Benson, {Relative permeability and trapping of CO 2 and water in sandstone rocks at reservoir conditions}, Water Resources Research 48~(2) (2012) 1--16.
\newblock \href {https://doi.org/10.1029/2011WR010859} {\path{doi:10.1029/2011WR010859}}.

\bibitem{Mehmani2018}
Y.~Mehmani, H.~A. Tchelepi, {Multiscale computation of pore-scale fluid dynamics: Single-phase flow}, Journal of Computational Physics 375 (2018) 1469--1487.
\newblock \href {https://doi.org/10.1016/j.jcp.2018.08.045} {\path{doi:10.1016/j.jcp.2018.08.045}}.

\bibitem{Mehmani2015}
A.~Mehmani, Y.~Mehmani, M.~Prodanovi{\'{c}}, M.~Balhoff, {A forward analysis on the applicability of tracer breakthrough profiles in revealing the pore structure of tight gas sandstone and carbonate rocks}, Water Resources Research 51~(6) (2015) 4751--4767.
\newblock \href {https://doi.org/10.1002/2015WR016948} {\path{doi:10.1002/2015WR016948}}.

\bibitem{Lautenschlaeger2019}
M.~P. Lautenschlaeger, H.~Hasse, Transport properties of the {Lennard-Jones} truncated and shifted fluid from non-equilibrium molecular dynamics simulations, Fluid Phase Equilibria 482 (2019) 38--47.
\newblock \href {https://doi.org/10.1016/j.fluid.2018.10.019} {\path{doi:10.1016/j.fluid.2018.10.019}}.

\bibitem{Lautenschlaeger2019a}
M.~P. Lautenschlaeger, H.~Hasse, Shear-rate dependence of thermodynamic properties of the {Lennard-Jones} truncated and shifted fluid by molecular dynamics simulations, Physics of Fluids 31~(6) (2019).
\newblock \href {https://doi.org/10.1063/1.5090489} {\path{doi:10.1063/1.5090489}}.

\bibitem{Diewald2020}
F.~Diewald, M.~P. Lautenschlaeger, S.~Stephan, K.~Langenbach, C.~Kuhn, S.~Seckler, H.-J. Bungartz, H.~Hasse, R.~M{\"{u}}ller, Molecular dynamics and phase field simulations of droplets on surfaces with wettability gradient, Computer Methods in Applied Mechanics and Engineering 361 (2020) 112773.
\newblock \href {https://doi.org/10.1016/j.cma.2019.112773} {\path{doi:10.1016/j.cma.2019.112773}}.

\bibitem{Lautenschlaeger2020}
M.~P. Lautenschlaeger, H.~Hasse, Thermal, caloric and transport properties of the {Lennard–Jones} truncated and shifted fluid in the adsorbed layers at dispersive solid walls, Molecular Physics 118~(9-10) (2020) e1669838.
\newblock \href {https://doi.org/10.1080/00268976.2019.1669838} {\path{doi:10.1080/00268976.2019.1669838}}.

\bibitem{Schmitt2022}
S.~Schmitt, T.~Vo, M.~P. Lautenschlaeger, S.~Stephan, H.~Hasse, Molecular dynamics simulation study of heat transfer across solid-fluid interfaces in a simple model system, Molecular Physics (2022) e2057364\href {https://doi.org/https://doi.org/10.1080/00268976.2022.2057364} {\path{doi:https://doi.org/10.1080/00268976.2022.2057364}}.

\bibitem{Huang2007}
H.~Huang, D.~T. Thorne, M.~G. Schaap, M.~C. Sukop, Proposed approximation for contact angles in {Shan-and-Chen-type} multicomponent multiphase lattice {B}oltzmann models, Physical Review E 76~(6) (2007) 1--6.
\newblock \href {https://doi.org/10.1103/PhysRevE.76.066701} {\path{doi:10.1103/PhysRevE.76.066701}}.

\bibitem{Latt2021}
J.~Latt, O.~Malaspinas, D.~Kontaxakis, A.~Parmigiani, D.~Lagrava, F.~Brogi, M.~B. Belgacem, Y.~Thorimbert, S.~Leclaire, S.~Li, F.~Marson, J.~Lemus, C.~Kotsalos, R.~Conradin, C.~Coreixas, R.~Petkantchin, F.~Raynaud, J.~Beny, B.~Chopard, {Palabos: parallel Lattice Boltzmann solver}, Computers {\&} Mathematics with Applications 81 (2021) 334--350.
\newblock \href {https://doi.org/10.1016/j.camwa.2020.03.022} {\path{doi:10.1016/j.camwa.2020.03.022}}.

\bibitem{Mu2019}
L.~Mu, X.~Liao, Z.~Chen, J.~Zou, H.~Chu, R.~Li, {Analytical solution of Buckley-Leverett equation for gas flooding including the effect of miscibility with constant-pressure boundary}, Energy Exploration {\&} Exploitation 37~(3) (2019) 960--991.
\newblock \href {https://doi.org/10.1177/0144598719842335} {\path{doi:10.1177/0144598719842335}}.

\bibitem{Washburn1921}
E.~W. Washburn, The dynamics of capillary flow, Physical Review 17~(3) (1921) 273--283.
\newblock \href {https://doi.org/10.1103/PhysRev.17.273} {\path{doi:10.1103/PhysRev.17.273}}.

\bibitem{Das2013}
S.~Das, S.~K. Mitra, {Different regimes in vertical capillary filling}, Physical Review E 87~(6) (2013) 063005.
\newblock \href {https://doi.org/10.1103/PhysRevE.87.063005} {\path{doi:10.1103/PhysRevE.87.063005}}.

\bibitem{Li2015}
K.~Li, D.~Zhang, H.~Bian, C.~Meng, Y.~Yang, {Criteria for Applying the Lucas-Washburn Law}, Scientific Reports 5~(1) (2015) 14085.
\newblock \href {https://doi.org/10.1038/srep14085} {\path{doi:10.1038/srep14085}}.

\bibitem{Das2012}
S.~Das, P.~R. Waghmare, S.~K. Mitra, {Early regimes of capillary filling}, Physical Review E 86~(6) (2012) 1--5.
\newblock \href {https://doi.org/10.1103/PhysRevE.86.067301} {\path{doi:10.1103/PhysRevE.86.067301}}.

\bibitem{Spaid1998}
M.~A. Spaid, F.~R. Phelan, {Modeling void formation dynamics in fibrous porous media with the lattice Boltzmann method}, Composites Part A: Applied Science and Manufacturing 29~(7) (1998) 749--755.
\newblock \href {https://doi.org/10.1016/S1359-835X(98)00031-1} {\path{doi:10.1016/S1359-835X(98)00031-1}}.

\bibitem{Silva2015}
G.~Silva, I.~Ginzburg, {The permeability and quality of velocity field in a square array of solid and permeable cylindrical obstacles with the TRT–LBM and FEM Brinkman schemes}, Comptes Rendus M{\'{e}}canique 343~(10-11) (2015) 545--558.
\newblock \href {https://doi.org/10.1016/j.crme.2015.05.003} {\path{doi:10.1016/j.crme.2015.05.003}}.

\bibitem{Phelan1996}
F.~R. Phelan, G.~Wise, {Analysis of transverse flow in aligned fibrous porous media}, Composites Part A: Applied Science and Manufacturing 27~(1) (1996) 25--34.
\newblock \href {https://doi.org/10.1016/1359-835X(95)00016-U} {\path{doi:10.1016/1359-835X(95)00016-U}}.

\bibitem{Parnas1995}
R.~S. Parnas, J.~G. Howard, T.~L. Luce, S.~G. Advani, {Permeability characterization. Part 1: A proposed standard reference fabric for permeability}, Polymer Composites 16~(6) (1995) 429--445.
\newblock \href {https://doi.org/10.1002/pc.750160602} {\path{doi:10.1002/pc.750160602}}.

\bibitem{Guo2002a}
Z.~Guo, C.~Zheng, B.~Shi, {Discrete lattice effects on the forcing term in the lattice Boltzmann method}, Physical Review E 65~(4) (2002) 046308.
\newblock \href {https://doi.org/10.1103/PhysRevE.65.046308} {\path{doi:10.1103/PhysRevE.65.046308}}.

\bibitem{Yu2010}
Z.~Yu, L.-S. Fan, {Multirelaxation-time interaction-potential-based lattice Boltzmann model for two-phase flow}, Physical Review E 82~(4) (2010) 046708.
\newblock \href {https://doi.org/10.1103/PhysRevE.82.046708} {\path{doi:10.1103/PhysRevE.82.046708}}.

\bibitem{Wood2015}
D.~L. Wood, J.~Li, C.~Daniel, Prospects for reducing the processing cost of lithium ion batteries, Journal of Power Sources 275 (2015) 234--242.
\newblock \href {https://doi.org/10.1016/j.jpowsour.2014.11.019} {\path{doi:10.1016/j.jpowsour.2014.11.019}}.

\bibitem{Weydanz2018}
W.~Weydanz, H.~Reisenweber, A.~Gottschalk, M.~Schulz, T.~Knoche, G.~Reinhart, M.~Masuch, J.~Franke, R.~Gilles, Visualization of electrolyte filling process and influence of vacuum during filling for hard case prismatic lithium ion cells by neutron imaging to optimize the production process, Journal of Power Sources 380 (2018) 126--134.
\newblock \href {https://doi.org/10.1016/j.jpowsour.2018.01.081} {\path{doi:10.1016/j.jpowsour.2018.01.081}}.

\bibitem{Westhoff2018}
D.~Westhoff, I.~Manke, V.~Schmidt, Generation of virtual lithium-ion battery electrode microstructures based on spatial stochastic modeling, Computational Materials Science 151 (2018) 53--64.
\newblock \href {https://doi.org/10.1016/j.commatsci.2018.04.060} {\path{doi:10.1016/j.commatsci.2018.04.060}}.

\bibitem{Lautenschlaeger2022}
M.~P. Lautenschlaeger, B.~Prifling, B.~Kellers, J.~Weinmiller, T.~Danner, V.~Schmidt, A.~Latz, Understanding electrolyte filling of lithium-ion battery electrodes on the pore scale using the lattice {B}oltzmann method, Batteries \& Supercaps (2022) e202200090\href {https://doi.org/https://doi.org/10.1002/batt.202200090} {\path{doi:https://doi.org/10.1002/batt.202200090}}.

\bibitem{Gostick2008}
J.~T. Gostick, M.~A. Ioannidis, M.~W. Fowler, M.~D. Pritzker, {Direct measurement of the capillary pressure characteristics of water-air-gas diffusion layer systems for PEM fuel cells}, Electrochemistry Communications 10~(10) (2008) 1520--1523.
\newblock \href {https://doi.org/10.1016/j.elecom.2008.08.008} {\path{doi:10.1016/j.elecom.2008.08.008}}.

\bibitem{Karpyn2009}
Z.~Karpyn, P.~Halleck, A.~Grader, {An experimental study of spontaneous imbibition in fractured sandstone with contrasting sedimentary layers}, Journal of Petroleum Science and Engineering 67~(1-2) (2009) 48--56.
\newblock \href {https://doi.org/10.1016/j.petrol.2009.02.014} {\path{doi:10.1016/j.petrol.2009.02.014}}.

\bibitem{Pini2012}
R.~Pini, S.~C. Krevor, S.~M. Benson, {Capillary pressure and heterogeneity for the CO2/water system in sandstone rocks at reservoir conditions}, Advances in Water Resources 38 (2012) 48--59.
\newblock \href {https://doi.org/10.1016/j.advwatres.2011.12.007} {\path{doi:10.1016/j.advwatres.2011.12.007}}.

\bibitem{Zhao2015}
H.~Zhao, Z.~Ning, Q.~Wang, R.~Zhang, T.~Zhao, T.~Niu, Y.~Zeng, {Petrophysical characterization of tight oil reservoirs using pressure-controlled porosimetry combined with rate-controlled porosimetry}, Fuel 154 (2015) 233--242.
\newblock \href {https://doi.org/10.1016/j.fuel.2015.03.085} {\path{doi:10.1016/j.fuel.2015.03.085}}.

\bibitem{Danner2016}
T.~Danner, S.~Eswara, V.~P. Schulz, A.~Latz, {Characterization of gas diffusion electrodes for metal-air batteries}, Journal of Power Sources 324 (2016) 646--656.
\newblock \href {https://doi.org/10.1016/j.jpowsour.2016.05.108} {\path{doi:10.1016/j.jpowsour.2016.05.108}}.

\bibitem{Tavangarrad2019}
A.~H. Tavangarrad, S.~M. Hassanizadeh, R.~Rosati, L.~Digirolamo, M.~T. van Genuchten, {Capillary pressure–saturation curves of thin hydrophilic fibrous layers: effects of overburden pressure, number of layers, and multiple imbibition–drainage cycles}, Textile Research Journal 89~(23-24) (2019) 4906--4915.
\newblock \href {https://doi.org/10.1177/0040517519844209} {\path{doi:10.1177/0040517519844209}}.

\bibitem{Zhu2021}
L.~Zhu, H.~Zhang, L.~Xiao, A.~Bazylak, X.~Gao, P.-C. Sui, {Pore-scale modeling of gas diffusion layers: Effects of compression on transport properties}, Journal of Power Sources 496~(June 2020) (2021) 229822.
\newblock \href {https://doi.org/10.1016/j.jpowsour.2021.229822} {\path{doi:10.1016/j.jpowsour.2021.229822}}.

\bibitem{Sauter2020}
C.~Sauter, R.~Zahn, V.~Wood, {Understanding Electrolyte Infilling of Lithium Ion Batteries}, Journal of The Electrochemical Society 167~(10) (2020) 100546.
\newblock \href {https://doi.org/10.1149/1945-7111/ab9bfd} {\path{doi:10.1149/1945-7111/ab9bfd}}.

\bibitem{Welge1952}
H.~J. Welge, {A Simplified Method for Computing Oil Recovery by Gas or Water Drive}, Journal of Petroleum Technology 4~(04) (1952) 91--98.
\newblock \href {https://doi.org/10.2118/124-G} {\path{doi:10.2118/124-G}}.

\bibitem{Jung2021}
H.~Jung, C.~Meile, {Pore-Scale Numerical Investigation of Evolving Porosity and Permeability Driven by Biofilm Growth}, Transport in Porous Media 139~(2) (2021) 203--221.
\newblock \href {https://doi.org/10.1007/s11242-021-01654-7} {\path{doi:10.1007/s11242-021-01654-7}}.

\bibitem{Shodiev2021}
A.~Shodiev, E.~Primo, O.~Arcelus, M.~Chouchane, M.~Osenberg, A.~Hilger, I.~Manke, J.~Li, A.~A. Franco, Insight on electrolyte infiltration of lithium ion battery electrodes by means of a new three-dimensional-resolved lattice {B}oltzmann model, Energy Storage Materials 38 (2021) 80--92.
\newblock \href {https://doi.org/10.1016/j.ensm.2021.02.029} {\path{doi:10.1016/j.ensm.2021.02.029}}.

\bibitem{Bhatnagar1954}
P.~L. Bhatnagar, E.~P. Gross, M.~Krook, A model for collision processes in gases. i. small amplitude processes in charged and neutral one-component systems, Physical Review 94~(3) (1954) 511--525.
\newblock \href {https://doi.org/10.1103/PhysRev.94.511} {\path{doi:10.1103/PhysRev.94.511}}.

\bibitem{Pope1980}
G.~A. Pope, {The Application of Fractional Flow Theory to Enhanced Oil Recovery}, Society of Petroleum Engineers Journal 20~(03) (1980) 191--205.
\newblock \href {https://doi.org/10.2118/7660-PA} {\path{doi:10.2118/7660-PA}}.

\end{thebibliography}


\pagebreak

\begin{center}
  \begin{singlespace}
  \Large Supporting Information: Homogenized Lattice Boltzmann Model for Simulating Multi-Phase Flows in Heterogeneous Porous Media\\[.5cm]
  \normalsize Martin P. Lautenschlaeger$^{a,b,*}$, Julius Weinmiller$^{a,b}$, Benjamin Kellers$^{a,b}$, Timo Danner$^{a,b}$, Arnulf Latz$^{a,b,c}$\\[.3cm]
  {\itshape \small ${}^a$German Aerospace Center (DLR), Institute of Engineering Thermodynamics, 70569 Stuttgart, Germany\\
  ${}^b$Helmholtz Institute Ulm for Electrochemical Energy Storage (HIU), 89081 Ulm, Germany\\
  ${}^c$Ulm University (UUlm), Institute of Electrochemistry, 89081 Ulm, Germany\\[.3cm]}
  \small ${}^*$Corresponding author: Martin.Lautenschlaeger@dlr.de\\[1cm]
  \end{singlespace}
\end{center}

\setcounter{section}{0}
\setcounter{equation}{0}
\setcounter{figure}{0}
\setcounter{table}{0}
\setcounter{page}{1}
\renewcommand{\thesection}{S\arabic{section}}
\renewcommand{\theequation}{(S\arabic{equation})}
\renewcommand{\thefigure}{S\arabic{figure}}
\renewcommand{\thetable}{S\arabic{table}}

In the present paper, the new homogenized multi-component Shan-Chen method (HMCSC) was presented and tested against typical benchmark cases for single- and two-phase flow in porous media. In addition to the validation in the paper, the bubble test in structurally resolved porous media as well as another interesting benchmark scenario, i.e.\ the Buckley-Leverett waterflooding experiment, are shown in the following. 

\clearpage

\section{Bubble Test in Structurally Resolved Idealized Porous Media} \label{sec:nonHomBubble}
The influence of the wettability of a porous medium on the bubble test discussed in \autoref{sec:BubbleTest} in the paper was validated by a scenario in which the porous medium was structurally resolved. The dimensions of the simulation domain along the $x$- and $y$-direction were $H=300$\,lu. The simulation setup was similar to the one shown in Figure~6 in the paper. Only the homogenized porous medium was replaced by 25 structurally resolved circular solid particles with a radius and an edge-to-edge distance of 15\,lu each. All lattice cells had the bounce-back fraction $n_\mathrm{s}=0$. Bounce-back conditions were applied at the solid surface and the wettability was varied in the range $\theta\approx\{60,90,120\}\degree$ by adjusting $G_{\mathrm{ads},1}=\{-0.2,0.0,0.2\}$ correspondingly (cf.\ \autoref{eq:contactAngle} in the paper). In addition, the initial bubble radius $R$ was varied in a range $R\approx[0.8,1.2]\,R_0$ where $R_0=106$\,lu.

\begin{figure}
	\centering
	  \includegraphics[width=1.0\textwidth]{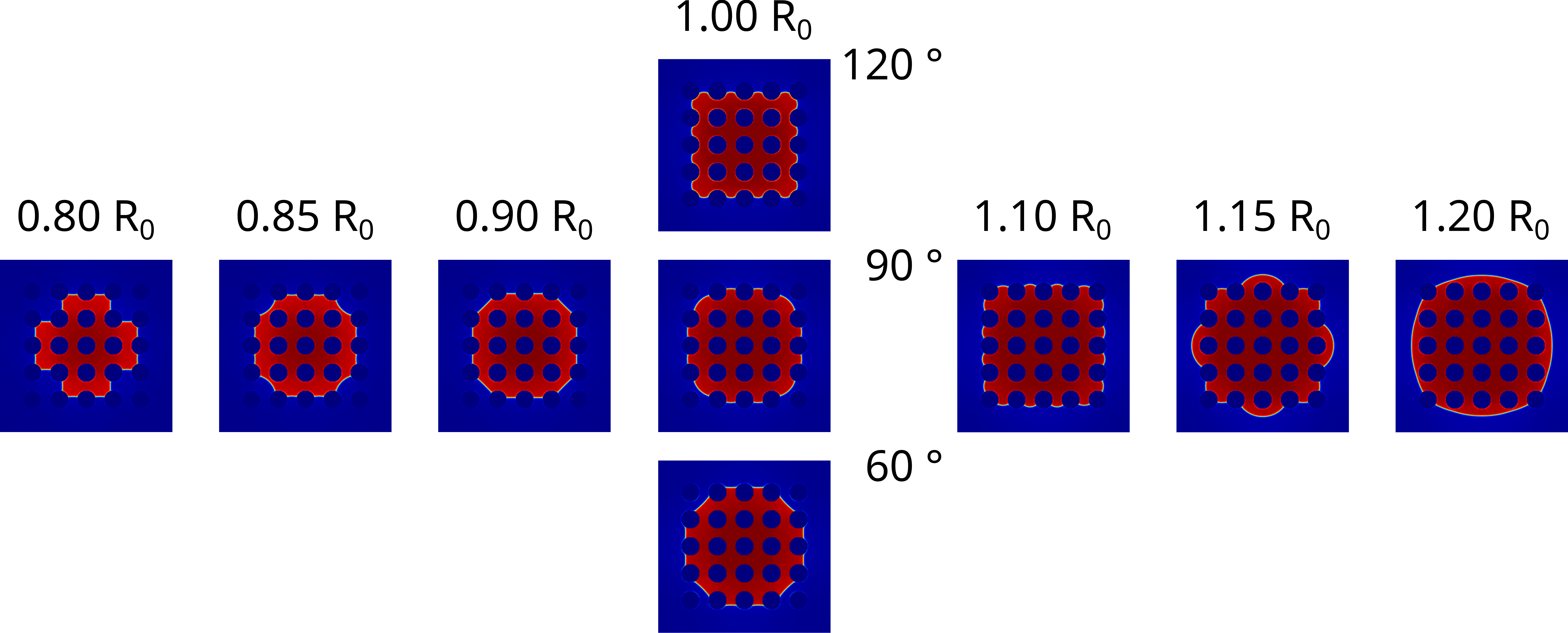}
	\caption{Bubble test in structurally resolved idealized porous media. Simulation results are shown for different wetting conditions and initial radii. }
	\label{fig:nonHomBubbleTest}
\end{figure}

The results are shown in \autoref{fig:nonHomBubbleTest}. There, the influence of the contact angle at constant $R=R_0$ is shown as vertical image sequence and the influence of $R$ at constant $\theta=90\degree$ is shown as horizontal image sequence. Two main points are observed: 1) In a coarse structurally resolved porous medium as it is used here, the shape of the bubble is strongly affected by both $R$ and $\theta$. However, note that the  finer the structure is compared to the bubble, i.e.\ with decreasing size of solid particles compared to $R$, the circularity of the bubble improves. Thus, converging towards a perfect circle or sphere in fine structures, emphasising the usefulness and accuracy of HMCSC especially for porous media where the characteristic length scale is significantly smaller than that of the flow. 2) Although the bubble shapes differ at constant initial radius $R=R_0$, the equivalent radius $R_\textrm{eq}=\sqrt{A/\pi}$ (where $A$ is the area covered by the bubble) at steady state is hardly affected by the wettability and varies in a range of $\pm$0.3\,\% only for the range of $\theta$ studied here. Thus, supporting the findings from \autoref{sec:BubbleTest} in the paper.

\section{Buckley-Leverett Experiment} \label{sec:BuckleyLeverett}
The Buckley-Leverett theory describes waterflooding, i.e.\ one fluid component displacing another fluid component from a porous medium. Under the following assumptions an analytical solution for this problem can be derived. Following Pope \cite{Pope1980} these are: 1) both fluids are immiscible and incompressible, 2) the porous medium is homogeneous and isotropic, 3) the flow is 1D, 4) the fluids are at local equilibrium, 5) gravity is negligible, 6) Darcy's law can be applied, and 7) the fluid injection is continuous and at constant composition. 

Based on the conservation of masses of both fluid components, the Buckley-Leverett equation for fluid 1 reads
\begin{align}    
    \label{eq:BuckleyLeverettEq}
    \phi \frac{\partial S_{\mathrm{1}}}{\partial t} + q \frac{\partial f}{\partial S_{\mathrm{1}}} \frac{\partial S_{\mathrm{1}}}{\partial x} = 0,
\end{align}
where $x$ and $t$ are position and time, respectively, $\phi$ is the porosity of the medium, $S_{\mathrm{1}}$ is the saturation of fluid 1, $q$ is its volumetric flux, and $f$ is the fractional-flow function. It is determined as
\begin{align}    
    \label{eq:FractionalFlowFunction}
    f = \left(1+\frac{k_{\mathrm{r,2}}}{k_{\mathrm{r,1}}}\right)^{-1},
\end{align}
where $k_{\mathrm{r},i}$ are the relative permeabilities of the fluids $i=\{1,2\}$. 

As was shown in \autoref{fig:Permeability}b) in the main paper, for the model parameters used for the validation, the relative permeabilities of both fluids show an almost linear dependence on the saturation. In such cases the relative permeability is typically described as
\begin{subequations}
\label{eq:BuckleyLeverettPerm}
    \begin{align}
      k_{\mathrm{r,1}} = S_{\mathrm{1}}^a, \label{eq:BuckleyLeverettPerm1}
    \end{align}
    \begin{align}
      k_{\mathrm{r,2}} = \left(1-S_{\mathrm{1}}\right)^a, \label{eq:BuckleyLeverettPerm2}
    \end{align}
\end{subequations}
where the parameter $a$ is typically chosen in the range of $a=]1,1.5]$ \cite{Mu2019}. This guarantees that the fractional-flow function $f$ has an inflection point (cf.\ \autoref{eq:FractionalFlowFunction}), which is a necessary requirement to construct the Buckley-Leverett solution. Therefore, it is assumed that $\mathrm{d} S_{\mathrm{1}}=0$ at the shock front between fluid 1 and fluid 2. Then, is velocity $v_{\mathrm{BL}}$ is determined from \autoref{eq:BuckleyLeverettEq} as
\begin{align}    
    \label{eq:BuckleyLeverettVel}
    v_{\mathrm{BL}} = \frac{q}{\phi} \frac{\partial f}{\partial S_{\mathrm{1}}},
\end{align}
where $\partial f/\partial S_{\mathrm{1}}$ can be interpreted as the tangent to $f$ and is determined as proposed by Welge \cite{Welge1952}.   

As all aforementioned assumptions can be met using LBM, the HMCSC is also tested for the Buckley-Leverett waterflooding experiment.

The simulation scenario consists of a homogenized porous medium in which all lattice cells have the same value for $n_s=n_{s,1}=n_{s,2}$ and $G_{\mathrm{ads},12}=0.0$. The system is 2D and its dimensions along the $x$- and $y$-direction are $H=100$\,lu and $L=5$\,lu, respectively. The system is initially filled with fluid 2. The densities of fluid 1 and fluid 2 are prescribed at the inlet $\left(\rho_1(x=0)\right)$ and at the outlet $\left(\rho_2(x=H)\right)$, respectively. Periodic boundary conditions are applied along the $y$-direction. Fluid 1 is injected through the inlet and the fluid flow is driven by applying the body force $G=10^{-4}$ in $+y$-direction. The influence of the solid fraction was studied in the range $n_s=\{0.1,0.5,0.9\}$.

\begin{figure}
	\centering
	  \includegraphics[width=0.8\textwidth]{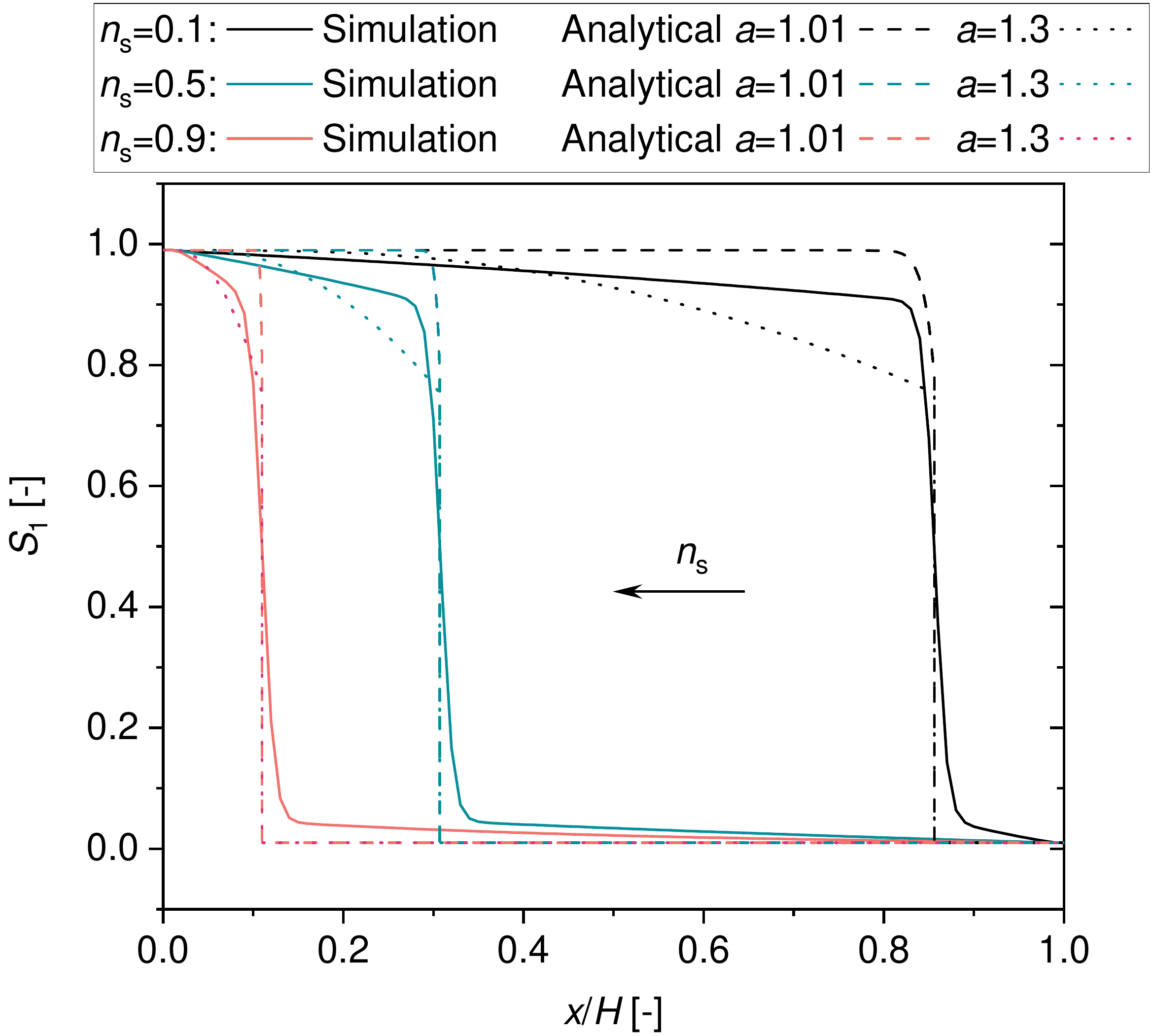}
	\caption{Profiles of the saturation of fluid 1 $S_{\mathrm{1}}$ at $t=200,000$\,ts for different solid fractions $n_s$. Simulation results determined using the HMCSC (solid lines) are compared with the analytical solutions of the Buckley-Leverett equation \autoref{eq:BuckleyLeverettEq} for the exponents $a=1.01$ and $a=1.3$ (cf.\ \autoref{eq:BuckleyLeverettPerm}).}
	\label{fig:BuckleyLeverett}
\end{figure}

\autoref{fig:BuckleyLeverett} shows the results of the Buckley-Leverett experiment at $t=200,000$\,ts for different values of $n_s$. The simulation results are in a reasonable overall agreement with the analytical solutions for $x\leq x_S$, where $x_S$ is the position of the shock front. The analytical solutions shows a strong dependence on parameter $a$ (cf.\ \autoref{eq:BuckleyLeverettPerm}). They can be interpreted as upper ($a=1.01$) and lower ($a=1.3$) bounds of the numerical solutions. Moreover, although the position of the shock front is predicted correctly, the shape of the shock front differs. While the analytical solutions shows a sharp change of $S_{\mathrm{1}}$, the simulation results show a diffuse interface. This is mainly due to numerical diffusion of the LBM and has also been observed in the literature \cite{Ning2019}. Nevertheless, the test shows that the HMCSC can be used to predict waterflooding experiments to some degree with limitations regarding the shape of the shock front.

\end{document}